%% file: ms2.tex
\documentclass[letterpaper]{article}

\usepackage{emulateapj}
\usepackage{onecolfloat}
\usepackage{apjfonts}
\usepackage{graphicx}

\newcommand{\citealp}{\cite}

\slugcomment{Received 2000 February 9; accepted 2000 April 18}

\shortauthors{Fossati et al. }
\shorttitle{Evolution of Mkn~421: II. Spectral Analysis and Phsyical Constraints}

\begin{document}
\twocolumn[
\title{X--ray Emission of Mkn~421: New Clues From Its Spectral Evolution:\\
II. Spectral Analysis and Physical Constraints}

\author{G. Fossati\altaffilmark{1,2}, 
A. Celotti\altaffilmark{3},
M. Chiaberge\altaffilmark{3},
Y. H. Zhang\altaffilmark{4},
L. Chiappetti\altaffilmark{4}, \\
G. Ghisellini\altaffilmark{5,6},
L. Maraschi\altaffilmark{5},
F. Tavecchio\altaffilmark{5},
E. Pian\altaffilmark{7},
A. Treves\altaffilmark{8} 
}

\begin{abstract}

Mkn~421 was repeatedly observed with \textit{Beppo}SAX in 1997--1998.
The source showed a very rich phenomenology, with remarkable spectral
variability.
This is the second of two papers presenting the results of a thorough
temporal and spectral analysis of all the data available to us, focusing in
particular on the flare of April 1998, which was simultaneously observed
also at TeV energies.
The spectral analysis and correlations are presented in this paper, while 
the data reduction and timing analysis are the content of the companion paper.
The spectral evolution during the flare has been followed over
few ks intervals, allowing us to detect for the first time the peak of
the synchrotron component shifting to higher energies during the
rising phase, and then receding.  
This spectral analysis nicely confirms the delay of the flare at the higher
energies, which in Paper~I we quantified as a hard lag of a few ks.
Furthermore, at the highest energies, evidence is found of variations of the
inverse Compton component.
The spectral and temporal information obtained challenge the simplest
models currently adopted for the (synchrotron) emission and most
importantly provide clues on the particle acceleration process.  
A scenario accounting for all the observational constraints is discussed,
where electrons are injected at progressively higher energies during the
development of the flare, and the achromatic decay is ascribed to the
source light crossing time exceeding the particle cooling timescales.
\end{abstract}

]
\altaffiltext{1}{Center for Astrophysics and Space Sciences, University of California at San Diego, 9500 Gilman Drive, La
Jolla, CA 92093-0424, U.S.A.}
\altaffiltext{2}{e--mail: gfossati@ucsd.edu}
\altaffiltext{3}{International School for Advanced Studies, via Beirut 2--4, 34014 Trieste, Italy}
\altaffiltext{4}{Istituto di Fisica Cosmica G.~Occhialini, via Bassini 15, 20133 Milano, Italy}
\altaffiltext{5}{Osservatorio Astronomico di Brera, via Brera 28, 20121 Milano, Italy}
\altaffiltext{6}{Osservatorio Astronomico di Brera, via Bianchi 46, 22055 Merate (Lecco), Italy}
\altaffiltext{7}{ITeSRE/CNR, via Gobetti 101, 40129 Bologna, Italy}
\altaffiltext{8}{Universit\`a dell'Insubria, via Lucini 3, 22100 Como, Italy}


\section{Introduction}
\label{sec:introduction}

Blazars are radio--loud AGNs characterized by strong variability,
large and variable polarization, and high luminosity.  
Radio spectra smoothly join the infrared-optical-UV ones.  
These properties are successfully interpreted in terms of synchrotron
radiation produced in relativistic jets and beamed into our direction due
to plasma moving relativistically close to the line of sight (e.g. Urry \&
Padovani \citealp{up95}).  
Many blazars are also strong and variable sources of GeV $\gamma$--rays,
and in a few objects, the spectrum extends up to TeV energies.  
The hard X to $\gamma$--ray radiation forms a separate spectral component,
with the luminosity peak located in the MeV--TeV range.
The emission up to X--rays is thought to be due to synchrotron
radiation from high energy electrons in the jet, while it is likely
that $\gamma$-rays derive from the same electrons via inverse Compton
(IC) scattering of soft (IR--UV) photons --synchrotron or ambient soft
photons (e.g. Sikora, Begelman \& Rees~\citealp{sbr94}; Ghisellini \& Madau
\citealp{gg_madau_96}; Ghisellini et al. \citealp{gg_sed98}).
The contributions of these two mechanisms characterize the average
blazar spectral energy distribution (SED), which typically shows two broad
peaks in a $\nu F_\nu$ representation (e.g. Fossati et
al. \citealp{fg_sed98}). 
The energies at which the peaks occur and their relative intensity provide
a powerful diagnostic tool to investigate the properties of the emitting
plasma, such as electron energies and magnetic field (e.g. Ghisellini et
al. \citealp{gg_sed98}).  
In X--ray bright BL Lacs (HBL, from High-energy-peak-BL Lacs, Padovani
\& Giommi \citealp{pg95}) the synchrotron maximum occurs in the
soft-X--ray band.

Variability studies constitute the most effective means to constrain the
emission mechanisms taking place in these sources as well as the 
geometry and modality of the energy dissipation. 

The quality and amount of X--ray data on the brightest sources start to
allow thorough temporal analysis as function of energy and the 
characterization of the
spectral evolution with good temporal resolution.

Mkn~421 is the brightest HBL at X--ray and UV wavelengths and thus it is
the best available target to study in detail the properties of the
variability of the highest frequency portion of the synchrotron component,
which traces the changes in the energy range of the electron distribution 
which is most
critically affected by the details of the acceleration and cooling processes.

This paper is the second of two, which present the uniform analysis of the
X--ray variability and spectral properties from \textit{Beppo}SAX
observations of Mkn~421 performed in 1997 and 1998.
In Paper~I (Fossati et al. \citealp{fossatiI}) we presented the data
reduction and the timing analysis of the data, which
revealed a remarkably complex phenomenology.
The study of the characteristics of the flux variability in different
energy bands shows that significant spectral variability is accompanying
the pronounced changes in brightness.
In particular, a more detailed analysis of the remarkable flare observed in
1998 revealed that: i) the medium energy X--rays lag the soft ones, ii) the
post--flare evolution is achromatic, and iii) the light curve is
symmetric in the softest X--ray band, and it becomes increasingly asymmetric
at higher energies, with the decay being progressively slower that the rise.

The general guidelines which we followed for the data reduction and
filtering are described in Paper~I (in particular in \S2 and \S3.2). 
Here we will only report on details of the treatment of the data 
specific to the spectral fitting.

The paper is organized as follows.  
In Sections~\S\ref{sec:bepposax_overview} and \S\ref{sec:observations} we
briefly summarize the basic information on \textit{Beppo}SAX and 
the 1997 and 1998
observations.
The results and discussion relative to the spectral analysis are the
content of \S\ref{sec:spectral_analysis} and \S\ref{sec:discussion}.  
In particular, the observed variability behavior strongly constrains any
possible time dependent particle acceleration prescription.
We will therefore consider these results together with those of the 
temporal 
analysis, discuss which constraints are provided to current models
and present
a possible scenario to interpret the complex spectral and temporal findings
(\S\ref{sec:interpretation}).  
Finally, we draw our conclusions in \S\ref{sec:conclusions}.

\section{\textit{Beppo}SAX overview}
\label{sec:bepposax_overview}

For an exhaustive description of the Italian/Dutch \textit{Beppo}SAX
mission we refer to Boella et al. (\citealp{boella97}) and references
therein.  
The results discussed in this paper are based on the data obtained with 
the Low and Medium Energy Concentrator Spectrometers (LECS and MECS)
and the Phoswich Detector System (PDS).  
The LECS and MECS have imaging capabilities in the 0.1--10~keV and
1.3--10~keV energy band, respectively, with energy resolution of 8\% at 6~keV.  
The PDS covers the range 13--300~keV.

The present analysis is based on the {\scshape SAXDAS} linearized
event files for the LECS and the MECS experiments, together with
appropriate background event files, as produced at the
\textit{Beppo}SAX Science Data Center ({\scshape rev~0.2, 1.1 and 2.0}).  
The PDS data reduction was performed using the XAS software
(Chiappetti \& Dal Fiume \citealp{chiappetti_dalfiume}) according to the
procedure described in Chiappetti et al. (\citealp{chiappetti_2155}).

\section{Observations}
\label{sec:observations}

\begin{figure}[t]
\centerline{\includegraphics[width=0.90\linewidth]{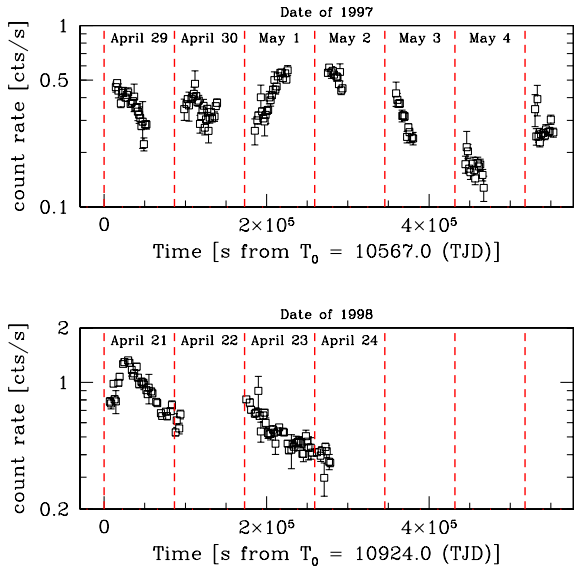}}
\vspace{-0.4cm}
\caption{\footnotesize\protect\baselineskip 8pt%
Light curves for the 1997 (top panel) and 1998 (bottom panel) campaigns,
for the 4--6~keV MECS band.  The binning time is 1500~s.
The reference time TJD=10567.0 corresponds to 1997--April--29:00:00 UT, 
and TJD=10924.0 corresponds to 1998--April--21:00:00 UT.
}
\label{fig:lc_97_and_98}
\end{figure}

Mkn~421 has been observed by \textit{Beppo}SAX in the springs of 1997 and 1998.
For reference, the \textit{Beppo}SAX light curves for the 4--6~keV band are
reported in Fig.~\ref{fig:lc_97_and_98}.  
Most of the forthcoming analysis is focused on the spectral
variability observed during the flare of 1998 April 21$^\mathrm{st}$,
clearly visible in the bottom panel.  
The journal of observations is given in Table~1 of Paper~I.

\section{Spectral Analysis}
\label{sec:spectral_analysis}

In order to unveil the spectral variability during and after the
flare, i.e. perform a time resolved spectral analysis, we subdivided
the whole dataset in sub--intervals corresponding to single
\textit{Beppo}SAX orbits (42 for the 1997 dataset, 16 on 1998 April
21$^\mathrm{st}$, 19 on 1998 April 23$^\mathrm{rd}$) or a grouping of them
sufficient to reach the same statistics (these are reported in
Table~\ref{tab:spectral_analysis_97}, together with the median time
[UTC] of each sub--interval).

LECS and MECS spectra have been accumulated as described in Paper~I.
LECS data have been considered only in the range 0.12--3~keV due to
calibration problems (a spurious hardening) at higher energies (Guainazzi
1997, private comm.)\footnote{A word of caution is necessary about
localized features probably related with calibration problems yet to be
solved (e.g. E$\sim$ 0.29~keV, $\sim$ 2 keV, in correspondence to the
Carbon edge and the Gold features due to the optics, respectively).}

The nominal full resolution spectra (i.e. channels \#11--285 for
0.1--3~keV in the LECS, and \#36--220 for 1.6--10~keV in the 
MECS) have been then rebinned using the grouping templates available at
\textit{Beppo}SAX--SDC%
\footnote{The energy resolution of the LECS and MECS detectors is
lower than the energy spacing between instrumental calibrated channels
($\sim$ 10~eV for LECS and 45~eV for MECS). 
Therefore it is compelling to rebin the spectra even in cases (like ours)
where the statistics in each low energy channel is good, in order to avoid
to overweight (as $\chi^2_\mathrm{a} + \chi^2_\mathrm{b} >
\chi^2_\mathrm{a+b}$) information that is not truly independent.
\textit{Beppo}SAX--SDC templates are downloadable at 
\url{\tt ftp://www.sdc.asi.it/pub/sax/cal/responses/grouping/}.
Due to the very steep spectral shape, in order to maintain a good
statistics in each new bin, we altered the templates for MECS above 
$\simeq 7$~keV, increasing the grouping of the original PI channels.}
The background has been evaluated from the blank fields provided by the
\textit{Beppo}SAX--SDC\footnote{%
Blank fields event files were accumulated on five different pointings of
empty fields and are available at the anonymous ftp: 
\url{\tt ftp://www.sdc.asi.it/pub/sax/cal/bgd/}.}, 
using an extraction region similar in size and position to the source
extraction region.  The Nov. 1998 release of public calibration files,
matrices and effective areas was used.

For the PDS spectra we applied the improved screening, implementing the
temperature and energy dependence of the pulse rise time (so--called PSA
correction method).

Due to a slight mis--match in the cross--calibration among the
different detectors, it has been necessary to include in the fitting
models multiplicative factors of the normalization (LECS/MECS and
PDS/MECS ratios).  The correct absolute flux normalization is provided
by the MECS (the agreement between MECS units is within the 2--3\%
limit of the systematics).  The expected value of these constant
factors is now well known and does not constitute a major/additional
source of uncertainty: according to Fiore, Guainazzi \& Grandi
(\citealp{cookbook}) the acceptable range for LECS/MECS is 0.7--1.0, with
some dependence on the source position in the detectors, while the
PDS/MECS ratio can be constrained between 0.77 and 0.95.  The latter
parameter is indeed crucial in sources like Mkn~421 where there is a
significant possibility that a different component arises in the PDS
range, whose recognition critically depends on the
capability/sensitivity to reject the hypothesis that the PDS counts
can be accounted for by the extrapolation of the LECS--MECS spectrum
(see Section~\ref{sec:spectral_analysis:pds}).

\input{Tab_Curved_Fits.txt}
\input{Tab_PL_Fits.txt}


\begin{figure}[t]
\centerline{\includegraphics[width=1.10\linewidth]{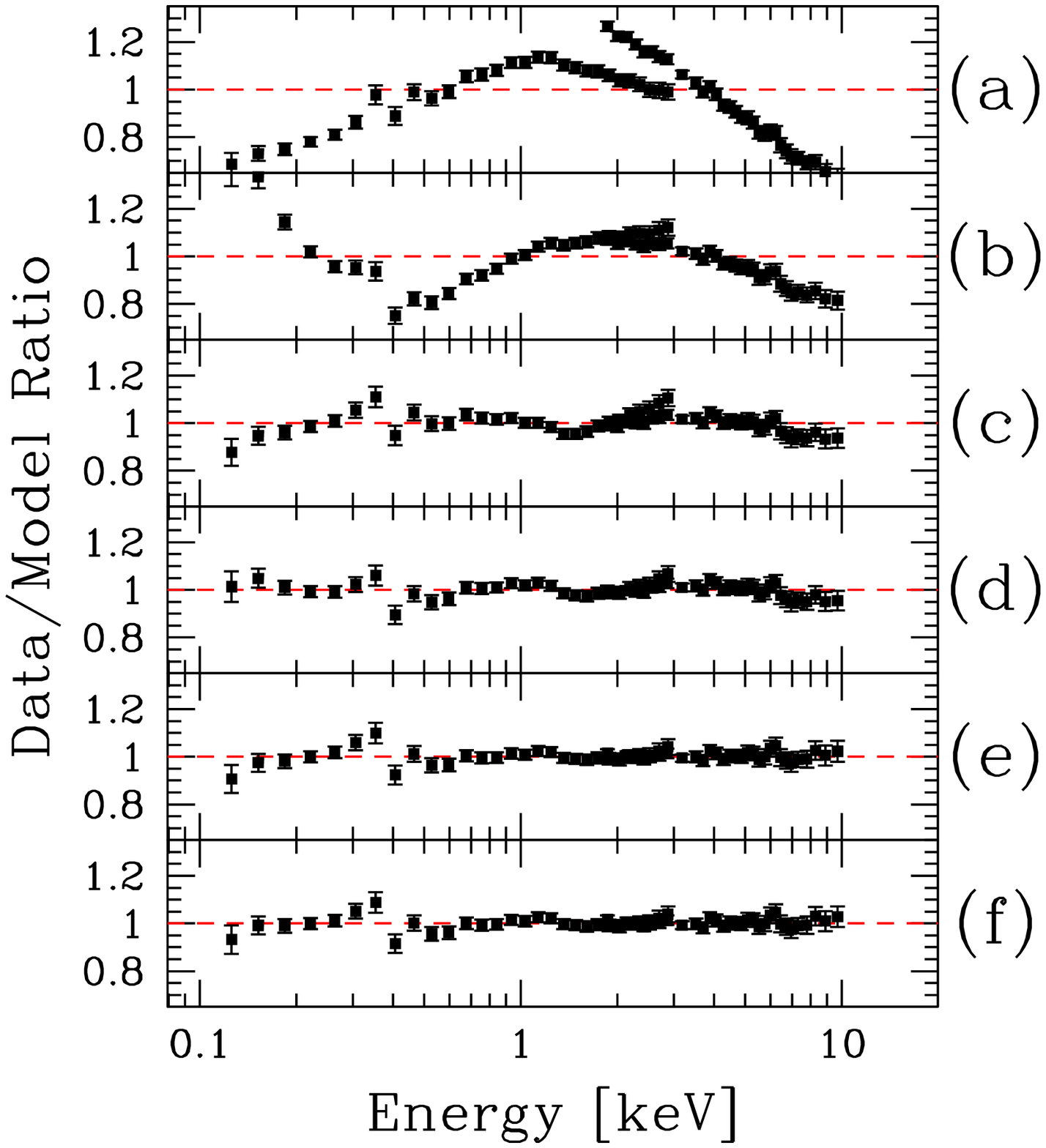}}
\vspace{-0.3cm} 
\caption{\footnotesize\protect\baselineskip 8pt%
Data/Model ratios for different spectral models applied to the 1998 data.
Top to bottom: single power law with (a) Galactic and (b) free
absorbing column, broken power law with (c) Galactic and (d) free
absorbing column, curved model, with (e) Galactic and (f) free
absorbing column.  The corresponding values of $\chi^2$ are reported
in Table~\ref{tab:power_laws_fit}.  } \label{fig:ratios_1998}
\end{figure} 

\subsection{Single and Broken Power Laws}
\label{sec:spectral_analysis:pl_bknpl}

We do not discuss in detail any power law spectral models.
For each LECS$+$MECS spectrum we fit the data with the single and the 
broken power law models\footnote{%
In Appendix~\ref{sec:appendix:bkn_pl_fits} we enclose a Table reporting the
best fit values for the the relevant parameters for the broken power law
model.}, both with free and fixed\footnote{%
In the direction of Mkn~421 the Galactic equivalent absorbing column is
N$_\mathrm{H} = (1.61 \pm 0.1) \times 10^{20}$ cm$^{-2}$ (Lockman \&
Savage~\citealp{lockman_savage95}), with the estimated uncertainty for high
Galactic latitudes.} 
absorbing column density.

Here we just show that they are an inadequate description of the downward
curved spectra\footnote{In Fossati et al. (\citealp{fg_lincei98}) we
presented a preliminary analysis of 1997 data, where LECS and MECS spectra
have been fit with a variety of combinations of single and broken power law
models to better describe the spectral curvature.  We refer the
interested reader to that work.}.
To illustrate the discrepancy, we took the data/model ratio for the best
fit continuum model, for each of the individual LECS$+$MECS spectra (16 for
1997, 12 for 1998), and we summed all of them, propagating the errors
accordingly.  
The results are shown in Figure~\ref{fig:ratios_1998} for the 1998 data
only, while in Table~\ref{tab:power_laws_fit} we report the total $\chi^2$
value for each trial model, together with the (simple) average value of
N$_\mathrm{H}$ for fits with free absorption.

Neither the single nor the broken power laws provide an adequate
representation of the data, not even when the value of the
absorbing column density is left free to vary\footnote{Although the plot
shown in Figure~\ref{fig:ratios_1998}d could appear as a good fit, the
significance of the small deviations above a few keV is indeed high,
yielding an unsatisfactory fit.} --a common workaround
used when the spectrum shows either a soft excess or a marked
steepening toward higher energies (e.g. Takahashi et
al.~\citealp{takahashi96}).

Indeed the description of the curvature by means of soft X--ray
absorption is not only un--physical, but it hinders the possibility of
extracting all the information from the data, by leaving as meaningful
parameter only the higher energy spectral index and accounting for all
the other effects by N$_\mathrm{H}$.  Furthermore this yields spectral
indices that are only rough estimates, as it is clear that the best
fit single power law does not describe the observed data in any (even
narrow) energy range (see Figure~\ref{fig:ratios_1998}a,b).
Finally, in the case of Mkn~421 there is no reason to postulate
any intrinsic absorbing component responsible
for the observed spectral curvature (see also
\S\ref{sec:spectral_analysis:absorption_features}).

\subsection{The Curved Model}
\label{sec:spectral_analysis:curved_model}

Motivated by the failure of the simplest power law models, we
developed a spectral model which is intrinsically curved\footnote{%
In Appendix~\ref{sec:appendix:curved_model} we discuss an important
caveat concerning the definition of curved models.},
with the aim of extracting all the information contained in the data,
and in particular estimate the position of the peak of the synchrotron
component, one of the crucial quantities in blazars modeling.

A spectral model providing an improved description of the continuum
shape also allows a more sensitive study of discrete spectral features
such as absorption edges, whose presence has been long searched in the
soft X--ray blazars spectra (e.g.  Canizares \& Kruper
\citealp{canizares84}; Sambruna et al. \citealp{sambruna_1426}; Sambruna \&
Mushotzky \citealp{sambruna_0548}).

We started from the following general description of a continuously
curved shape (see e.g. Inoue \& Takahara \citealp{inoue_takahara96}, 
Tavecchio, Maraschi \& Ghisellini \citealp{tavecchio_etal98}):
\begin{equation}
F(E) \;=\; K\; E^{-\alpha_{-\infty}}\; \left( 1 + \left(\frac{E}{E_{\rm
B}}\right)^f \right)^\frac{\alpha_{-\infty}-\alpha_{+\infty}}{f} 
\end{equation}
\noindent
where $\alpha_{-\infty}$ and $\alpha_{+\infty}$ are the asymptotic values 
of spectral indices for $E \ll E_\mathrm{B}$ and $E \gg E_\mathrm{B}$
respectively, while $E_\mathrm{B}$ and $f$ determine the scale length of
the curvature.

We re--expressed this function in terms of the spectral indices at
finite values of $E$, which characterize the local shape of the
spectrum, instead of the asymptotic ones, which do not have any direct
reference to the observed portion of the spectrum.

The spectral model is then expressed in a form such that the available
parameters are $(E_1, \alpha_1, E_2, \alpha_2, E_\mathrm{B}, f)$
instead of ($\alpha_{-\infty}$, $\alpha_{+\infty}$, $E_\mathrm{B}$,
$f$) (for more details see Appendix~\ref{sec:appendix:this_curved_model}).
As we have two extra parameters, for a meaningful use of this
spectral description we have to fix one for each of the pairs $(E_1,
\alpha_1)$ and $(E_2, \alpha_2)$.  
Eventually this degeneracy turns out to be a powerful property of this
model, because it allows us to derive the spectral index at selected
energies (setting $E_\mathrm{i}$ at the preferred values), or even more
interestingly to estimate the energy at which a certain spectral index is
obtained (setting $\alpha_\mathrm{i}$ at the desired value).  
The most relevant example of this latter possibility is the determination of
the position of the peak (as seen in $\nu F_\nu$ representation) of
the synchrotron component $E_\mathrm{peak}$ --if it falls within the
observed energy band-- and the estimate of the associated error.
This can be obtained by setting one spectral index, i.e.  $\alpha_1=1$, and
leaving the corresponding energy $E_1$ free to vary in the fit: 
the best fit value of $E(\alpha=1)$ gives $E_\mathrm{peak}$.

\begin{figure}[t]
\centerline{\includegraphics[width=1.00\linewidth]{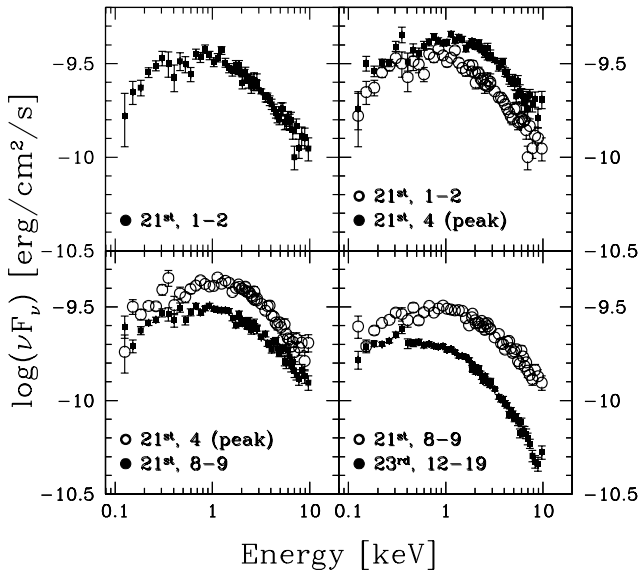}}
\vspace{-0.6cm}
\caption{\footnotesize\protect\baselineskip 8pt%
Inferred \textit{Beppo}SAX $\nu F_\nu$ spectra of Mkn~421 in 1998.
The plotted spectra correspond to the orbit groupings 1$+$2, 4, 8$+$9 of April
21$^\mathrm{st}$, and 12--19 of April 23$^\mathrm{rd}$ (see
Table~\ref{tab:spectral_analysis_98}).
For comparison we also plotted in light grey color the spectrum
of the previous box.}
\label{fig:quattro_xseds}
\end{figure}

\subsection{Results}
\label{sec:spectral_analysis:results}

Most of the spectral analysis with the curved model has been performed
by keeping the absorbing column fixed to the Galactic value, to remove
the degeneracy between the effects of a variable absorption and
intrinsic spectral curvature.  We will briefly comment on the fits
with a variable absorbing column in the next section.

In order to determine the values for the really interesting parameters, we
tried a set of values for the parameter $f$, ranging from 0.5 to 3 (with a
step of 0.5).  
All the other parameters were left free to vary during these step--fits.
We then selected the value of $f$ yielding the minimum total $\chi^2$.  
The best fit value for 1997 is $f = 1$, while for 1998 $f = 2$, as expected
from the stronger curvature of the spectra.  
In Table~\ref{tab:spectral_analysis_98} we report the spectral parameters,
together with the 0.2--1, 2--10 and 0.1--10~keV fluxes (de--reddened) and
$\chi^2$ values, for each of the time sliced spectra. 
The global Data/Model ratio is shown in Figure~\ref{fig:ratios_1998}e, and
the total $\chi^2$ is reported in Table~\ref{tab:power_laws_fit}.  
The curved model is strongly preferred also from the purely statistical
standpoint: in fact the model with Galactic N$_\mathrm{H}$ has a
significantly better $\chi^2$ than the broken power law with free
N$_\mathrm{H}$, although it has one less adjustable parameter (i.e. one
more degree of freedom).

Each spectrum has been fitted a few times in order to derive spectral
indices at several energies (0.5, 1, 5, 10~keV), and also
E$_\mathrm{peak}$. 
For consistency we checked each time that not only the value of the
$\chi^2$ remained the same, but also all the other untouched parameters
took the same value and confidence intervals.  The fits are indeed very
robust in this sense.

The main new result is that we were able to \textit{determine the energy of
the peak of the synchrotron component}, and to assign an error to it.
This has been possible with reasonable accuracy for both 1997 and 1998 spectra,
with a couple of cases yielding only an upper limit for the
peak energy\footnote{%
These are the cases ``35--37'' and ``41--42'' of 1997.
Actually the formal fitting yields a confidence interval, but with the
lower extreme falling outside of the observed energy range.}. 

An example of the remarkable spectral evolution during the 1998 flare
is shown by the deconvolved spectra in Figure~\ref{fig:quattro_xseds},
which also illustrates the strong convexity of the spectrum and the
well defined peak.  

In 1997 the source was in a lower brightness state, with an average X--ray
spectrum softer (at all energies $\Delta\alpha_{97\longrightarrow 98}
\simeq 0.4$), and a peak energy 0.5~keV lower.

\begin{figure}[t]
\centerline{\includegraphics[width=0.80\linewidth]{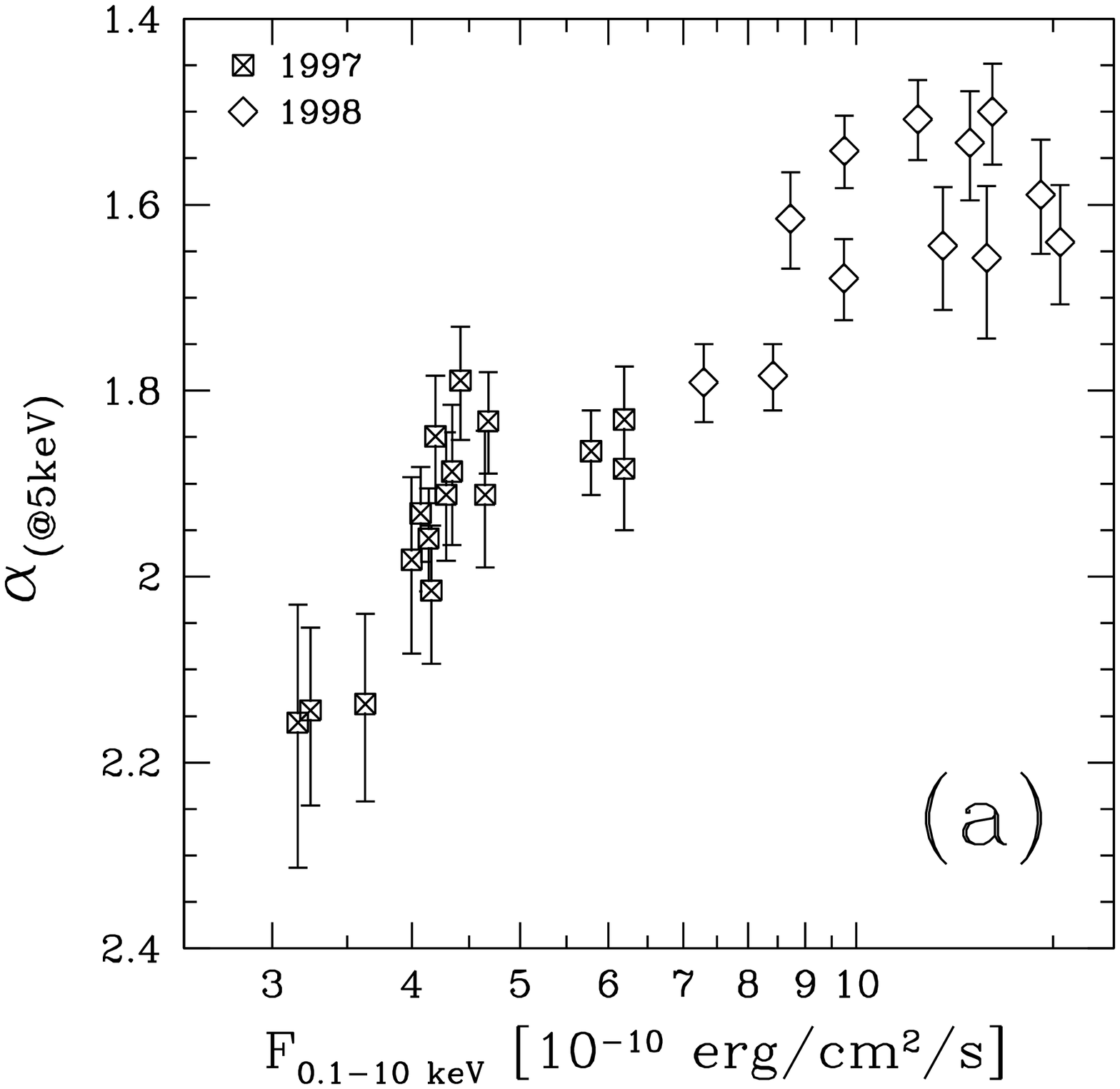}}
\vspace{-0.7cm}
\centerline{\includegraphics[width=0.80\linewidth]{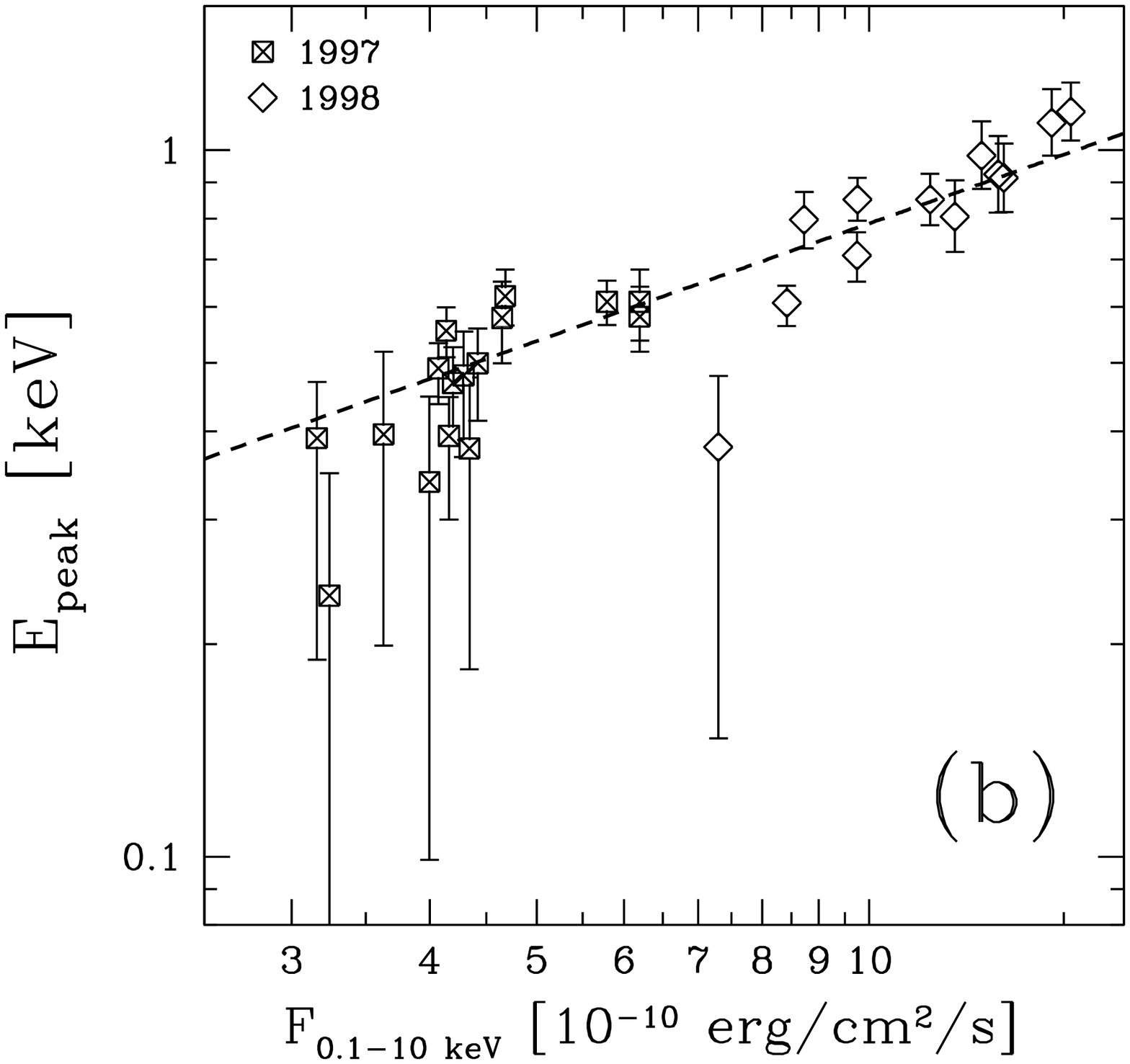}}
\vspace{-0.4cm}
\caption{\footnotesize\protect\baselineskip 8pt%
(a) spectral index at 5~keV and (b) synchrotron peak energy plotted versus
the de--absorbed 0.1--10~keV flux, both for 1997 and 1998 datasets.
The dashed line across the diagram (b) represents the best fitting power
law, with a slope $\epsilon=0.55$.
}
\label{fig:a5_vs_Flux}
\label{fig:epeak_vs_Flux}
\end{figure}

More globally, the analysis shows that there is a clear relation
between the flux variability and the spectral parameters, both in 1998
and, albeit less strikingly, in 1997.  Particularly important is the
correlation between changes in the brightness (even the small ones)
and shifts of the peak position, as the latter carries direct
information on the source physical properties.  
In Figure~\ref{fig:epeak_vs_Flux}a,b, $\alpha$ at 5~keV and
E$_\mathrm{peak}$ for 1997 and 1998 are plotted versus the 0.1--10~keV
flux.  The source reveals a coherent spectral behavior between 1997 and
1998 and through a large flux variability (a factor 5 in the 0.1--10~keV
band).  
Both the peak energies and the spectral indices show a tight relation with
flux, the latter being described by E$_\mathrm{peak} \propto
\mathrm{F}^\epsilon$, with $\epsilon = 0.55 \pm 0.05$.

\subsubsection{Hard Lag in 1998 spectra}
\label{sec:spectral_analysis:hard_lag}

\noindent 
One further important finding of the time resolved spectral study is the
signature of the hard lag which strengthens what already inferred from
the timing analysis (see Paper~I). 
In fact the local spectral slope at different energies indicates that the
flare starts in soft X--rays and then extends to higher energies.

\begin{figure}[t]
\centerline{\includegraphics[width=0.90\linewidth]{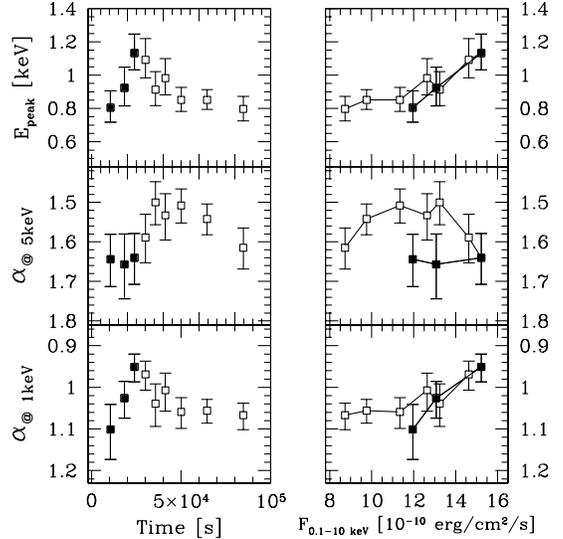}}
\vspace{-0.4cm}
\caption{\footnotesize\protect\baselineskip 8pt%
Top to bottom: the energy of the peak of the synchrotron component,
and the photon spectral indices at 5 and at 1~keV versus time (left
column) and de--absorbed 0.1--10~keV flux (right column).  The solid
black symbols represent the behavior before the top of the flare,
while empty dots represent the spectral evolution during the decay.  
The data points are connected to show the time sequence.
A shift of the synchrotron peak and a delay in the response of the
spectrum at 5~keV can be clearly seen.
}
\label{fig:dettaglio_98}
\end{figure}

A blow up of the 1998 flare interval is shown in
Figure~\ref{fig:dettaglio_98}: spectral indices at 1 and 5~keV, and
E$_\mathrm{peak}$ are plotted versus time in the left hand plots, and
versus flux in the right hand ones.  The synchrotron peak shifts
toward higher energies during the rise and then decreases as soon as the
flare is over.  
The spectral index at 1~keV reflects exactly the same behavior, as expected
being computed at the energy around which the peak is moving (and in fact
it moves in a narrow range around $\alpha=1$).
On the contrary, the spectral shape at 5~keV does not vary until a few ks
after the peak, and only then --while the flux is decaying and the peak is
already receding-- there is a response with a significant hardening of the
spectrum.  
The right side diagrams illustrate this behavior directly in terms of flux:
the evolution of the peak energy and the 1~keV spectral shape follow
closely that of the flare, while at 5~keV the spectrum does not change
shape until after the peak.

Thus the spectral evolution at higher energies develops during the
decay phase of the flare, tracing a loop in the $\alpha$ vs. Flux
diagram, with a significant hardening of the spectrum.  Due to the
presence of the hard lag, the loop is traced in the \textit{opposite}
way with respect the evolution that is commonly observed during flares in
HBL, that is a spectral hardening during the rise of the outburst, followed
by a softening during the decay (see however Sembay et al.
\citealp{sembay_2155}, Catanese \& Sambruna \citealp{catanese_sambruna_00}). 

This behavior --usually accompanied by a soft lag-- produces a clockwise
loop in $\alpha$ vs. Flux.

\subsubsection{Absorption Features: N$_{\rm H}$ and edges}
\label{sec:spectral_analysis:absorption_features}

We also fit the 1997 and 1998 spectra with the curved spectral model
leaving the N$_\mathrm{H}$ free to vary.  The only constraint that we
imposed is on the exponent $f$, which is set at the best fit values of
$f=1$ and $f=2$ for 1997 and 1998, respectively.  
The improvement of the global fit\footnote{%
$\Delta\chi^2 = 17.9$ for 16 additional \textit{d.o.f.} for 1997 and
$\Delta\chi^2 = 15.9$ for 12 additional \textit{d.o.f.} for 1998,
corresponding to an average $\Delta \chi^2 \simeq 1.2$ per spectrum for a
change in \textit{d.o.f.} from 57 to 56.} is not statistically significant. 
The average best fit absorbing column is indeed very close to and
consistent with the Galactic one, leading us to conclude that there is no
requirement whatsoever for an additional absorbing column to model the data.

We searched for the presence of absorption edges at low energies (E
$\lesssim$ 1~keV), both in 1997 and 1998 data.  We added to the curved
continuum an absorption edge, with a first guess fit at E$\simeq 0.4 -
0.5$~keV, as even in the case of the best curved model with free
N$_\mathrm{H}$ there are residues around this energy (see
Figure~\ref{fig:ratios_1998}f).  The most delicate issue is the
calibration problem related with the instrumental Carbon absorption
edge at 0.29~keV.  Even the smallest uncertainties ($\sim 1-2$\%) in
the knowledge of the instrument response around this energy can induce
significant spurious features, typically at slightly higher energy
(see Fossati \& Haardt \citealp{fg_fh_lecs}).

There is not convincing evidence for discrete absorption features.
In 5/28 cases the fit routine has not found an edge, while there are
indeed a few spectra (8/28) in which the fit definitely put the edge
at 0.27--0.30~keV, to account for systematic deviation due to
calibration uncertainties.  There are then 15/28 cases where the edge
energy is around 0.4--0.6~keV, but again only in a handful of cases
the best fit energy E$_\mathrm{edge}$ is significantly different from
0.29~keV.  

Summarizing, there are only three individual spectra (out of 28) for
which the inclusion of the edge is formally statistically significant
($>$ 95\%), according to the F--test.  
These three cases do not stand out in the sample for any other property,
such as brightness or curvature of the continuum.
Best fit energies and optical depths are typically E$_\mathrm{edge}\simeq
500^{+50}_{-100}$~eV, and $\tau \simeq 0.4^{+0.3 - 0.6}_{-0.2}$.  
However, considering that the sample comprises 28 spectra, with the
significance threshold set at 95\%, three positive detections can not be
regarded as a compelling evidence for the presence of absorption edges.  

\subsubsection{PDS data}
\label{sec:spectral_analysis:pds}

Covering the range above $\simeq 12$~keV, the PDS instrument could
detect the IC component, which might start to dominate the emission in
this band.  It has always proven to be very difficult to
detect/constrain this component, although it is expected to have a
very hard spectrum, that should make it easier to disentangle from the
very steep tail of the synchrotron component.  There are a few
objects, the so--called intermediate BL Lacs, in which the cross--over
between synchrotron and IC occurs in the 0.1--10~keV band, for which
it has then been possible to clearly detect the hard IC power law,
with typical $\alpha \simeq 0.3-0.7$ (e.g. Tagliaferri et
al. \citealp{tagliaferri_on231} for ON~231; Giommi et
al. \citealp{giommi_0716} for S5~0716$+$714;
Sambruna et al. \citealp{sambruna_bllac_99} for BL Lac).
The good sensitivity of the PDS yielded a convincing evidence of IC
emission in the case of PKS~2155$-$304 (Giommi et al.
\citealp{giommi_2155}).  The preliminary analysis of the Mkn~421 1997
data (Fossati et al. \citealp{fg_lincei98}) suggested the presence of a
hard component.  We therefore focused on the search for the IC
component for both of 1997 and 1998 PDS datasets.

For 1998, when the source was brighter, we accumulated PDS spectra
according to the same partition of the light curves used for the
0.1--10~keV data.  In most of the individual spectra there is a
significant ($\ge$3-$\sigma$) detection only up to 40~keV.  The light
curve for the 12--26~keV band is shown in Figure~3 of Paper~I.

We then restricted the analysis to only one spectrum for 1997,
integrating over the whole campaign, and two for 1998, one for April
21$^\mathrm{st}$, and one for April 23$^\mathrm{rd}$.  The integration
times are 54.8, 21.1 and 25.2 ks respectively (see Paper~I).  
We grouped the data in 6 channels spectra with boundaries at 12, 18, 27,
40, 60, 90 and 130 keV.

In the 1997 spectrum there is a positive detection up to 90~keV.  For
April 21$^\mathrm{st}$ --corresponding to the synchrotron flare--
there is a strong signal up to 60~keV, while, surprisingly, in the
April 23$^\mathrm{rd}$ spectrum there is a $\gtrsim$3-$\sigma$
detection in each individual bin up to 130~keV.  From 12 to 40 keV the
April 21$^\mathrm{st}$ count rates are significantly higher than those
of April 23$^\mathrm{rd}$, but while the former continue to steeply
decrease with increasing energy, the latter flatten above 40~keV, with
a cross over in the 40--60~keV bin.  As the integration times for
these two spectra are similar (and in fact statistical uncertainties
on the count rate are of the same order), this cannot account for this
difference\footnote{%
Also, we are not aware of any intra--observation ``background
fluctuations'' to be taken into account in the standard PDS data reduction
(which has been performed by following carefully the prescriptions given by
the instrument team, for all the details refer to Chiappetti et al.
\citealp{chiappetti_2155}), nor of any long term ``background
fluctuations'' that could be responsible for the observed variation.
No such thing was reported neither by the PDS instrument team, nor by the
\textit{Beppo}SAX Science Data Center.}.

The results of single power law fits are reported in
Table~\ref{tab:pds_spectra}: clearly the April 23$^\mathrm{rd}$
spectrum is significantly harder.  Moreover, the spectral indices for
the first two cases are consistent with those derived from the LECS
and MECS datasets and the general picture of a continuously
steepening spectrum, while this does not hold for the third case.

While this is already interesting, there is one additional, 
somehow unexpected, finding: apparently the April 23$^\mathrm{rd}$
harder spectrum cannot be simply interpreted as the detection of the
IC component when the synchrotron tail recedes. In fact a hard power law at
this flux level --if present-- would have been detected also on
April 21$^\mathrm{st}$, while the count rates in the higher energy
bins for April 21$^\mathrm{st}$ set a tight upper limit. This
behavior might be ascribed to either a different independent origin
or a delayed response of the IC component with respect to the
synchrotron one (indeed some evidence of a possible delayed decay
of the PDS with respect to the MECS emission can be seen in the 
light curves). 

\input{Tab_PDS_fits1.txt}
\input{Tab_PDS_fits2.txt}

Finally we tried to quantify the presence of the IC power law in the three
datasets by fitting MECS and PDS data together.
We proceeded in the following way:
\begin{itemize}
\item We accumulated LECS and MECS spectra over the same intervals of the
PDS ones, i.e. one cumulative spectrum for each day of 1998, and one for
the whole 1997.
\item We fit the LECS$+$MECS spectra in order to constrain the average
synchrotron spectrum.
\item We then fixed all the parameters of the curved model to the
LECS$+$MECS best fit values, and added to the model a hard power law
component\footnote{%
We used the {\tt pegpwrlw} (power law with pegged normalization) model of
{\tt XSPEC} that provides a robust measure of the normalization. 
In fact this is defined over a selectable energy band, instead of the
monochromatic value at 1~keV that can be strongly correlated with other
parameters if 1~keV does not fall close to the logarithmic median of the
band spanned by the data.} with fixed spectral index (we tried the values
$\alpha_\mathrm{IC} = 0.4, 0.5, 0.6, 0.7$).  
We also added an exponential cut--off to the synchrotron component setting
its two parameters, i.e. cut--off and e--folding energies, to 10 and 40
times the best fit value of E$_\mathrm{peak}$.
\item We used this partially constrained model to fit the MECS (above E$\ge
5$~keV) and PDS spectra. 
The two 1998 datasets have been fit jointly with PDS/MECS intercalibration
factor free to vary, but set to be the same for both, and of course
independent normalization.
\end{itemize}

\noindent
The results are reported in Table~\ref{tab:ic_pl_fits}, for the
$\alpha_\mathrm{IC} = 0.5$ case (the values relative to the other
$\alpha_\mathrm{IC}$ are not significantly different).  The best fit
PDS/MECS normalization ratio is 0.88.

There seems to be a significant change in the flux level of the
expected IC component between 1998 April 21$^\mathrm{st}$ and
23$^\mathrm{rd}$.  The best fit value for the normalization for the
first day is zero, while for the second day a non--zero IC flux is
definitely required.  In order to test the likelihood of the apparent
variation, we fit the data constraining the normalization to be the
same for the two datasets (third case in Table~\ref{tab:ic_pl_fits})
and the F--test probability to obtain by chance the observed change in
$\chi^2$ is $\sim$~0.01 ($\triangle \chi^2 = 6.08$, \textit{d.o.f.}
from 30 to 31).  

The basic result for 1997 is that the IC component is definitely detected.
Its brightness level is comparable with the average of 1998, while the
synchrotron component shows a significant variation.  

It should be noted that the synchrotron fluxes reported in
Table~\ref{tab:ic_pl_fits} are not the best reference because at those
energies, well above the synchrotron peak, the spectrum is very steep
and even the smallest shift of the peak energy translates in a big
change in the flux.  Nevertheless even when considering the energy
range 0.1--10~keV, which includes the synchrotron peak, the variation
between 1997 and 1998 is of the order of a factor of 2--3
(e.g. Figure~\ref{fig:dettaglio_98}).  The sampling of the 1997 light
curve is not good enough to enable us to investigate any
possible variation of the PDS spectrum with the source brightness.


\section{Discussion}
\label{sec:discussion}

Before focusing on the modeling of the temporal and spectral behavior
illustrated in this paper and in Paper~I, let us consider a few issues
which we consider of particular relevance for the interpretation of the
origin of blazar variability.


\subsection{Spectral Variability, Synchrotron Peak and IC component}
\label{sec:discussion:spectral}
\label{sec:epeak}

Soft and hard X--ray bands show a different behavior.
This might be attributed to the contribution from both the synchrotron and
inverse Compton components at these energies. 
As electrons emitting X--rays through the two
processes would have different energies, the two components are not
expected to vary on the same timescales and in phase.  Indeed on one
hand the constraints provided by the analysis of the PDS data suggest
that the synchrotron component constitutes the dominant contribution
to the flux in the 12--26~keV band on 1998 April 21$^\mathrm{st}$
(during the flare) with a light curve somewhat different from that
of the softer synchrotron X--rays.  On the other hand, the evidence
supporting the possibility of a substantial variation in the IC
component during the 1998 observations is extremely interesting.  We
can rule out the possibility that the flux level of the hard component
is approximately constant between the two observations of 1998 (with a
99\% confidence), and thus that its detectability only depends on the
the brightness level of the tail of the synchrotron emission. 

\subsection{The Flaring + Steady Components Hypothesis}

A further interesting issue raised in Paper~I in connection with the
temporal analysis is the presence and role of quasi--stationary emission. 
Our results support the view that the short--timescale, large--amplitude
variability events could be attributed to the development of new
individual flaring components, giving rise to a spectrum out--shining a
more slowly variable contribution.

Indeed there is growing evidence that the overall blazar emission
comprises a component that is (possibly) changing only on long
timescales and that does not take part in the flare events (see for
instance Mkn~501, Pian et al. \citealp{pian_mkn501_98}, and Pian et al.
\citealp{pian_3c279} for the case of the UV variability of 3C~279).

The deconvolution of the SED into different contributions would allow
the study of the evolution of the flaring component possibly clarifying
the nature of the dissipation events occurring in relativistic jets
--in particular the particle acceleration mechanism-- and the modality
and temporal characteristics of the initial release of plasma and
energy through these collimated structures.

The deconvolution is however still difficult to achieve with the
available temporal and spectral information.  
In paper~I we discussed the results of our attempt to measure directly the
relative contributions of what we identify as steady and flaring
components (parameter ${\cal R}$), from the characteristics of the flare
decay.  
The result is very interesting, although it does not provide a useful
handle on the evolving spectral properties of the flare itself.  
A more feasible approach is to assume specific forms of the flare evolution
and test them against the data. One of the simplest working hypothesis is to
assume that the flare evolution can be reproduced by the time
dependent flux and energy shifts of a peaked spectral component with fixed
spectral shape (e.g. Krawczynski et al. \citealp{krawczynski_mkn501_2000}).
In this context it is relevant to remind that the maximum of the
synchrotron emission corresponds to the energy at which most of the
energy in particles is located.  Its value and temporal behavior thus
trace the evolution of the bulk of the energy deposition into
particles by the dissipation mechanism.

It is therefore possible that the variability characteristics (both
temporal and spectral) might chiefly depend on the (particles/photons)
energy \textit{relative} to the synchrotron peak.  In particular, the
results on T$_\mathrm{short}$ (Paper~I) can be either related to the
dominance of the light crossing time over the cooling timescales, or
alternatively to the different position of the sampled energies with
respect to the synchrotron peak as above the peak the timescales are
shorter. In any case the fact that timescales are similar at the
different energies possibly indicates the evolution of a (observed)
fixed--shape flaring component.

\subsection{Sign of the lag}
\label{sec:sign_of_lag}

One of the main new results of the temporal analysis presented in Paper~I
is the significant detection of a hard lag, opposite to the behavior
commonly detected in HBL. Interestingly, a recent study by Zhang et
al. (\citealp{zhang_2155}) on the source PKS~2155--304 has shown the
presence of an apparent inverse trend between the source brightness
level and the duration of the (soft) lag. And intriguingly, the 1998
flare represents the brightest state ever observed from Mkn~421, thus
possibly suggesting that an extrapolation of the above phenomenological
trend might even account for a hard lag.  
This possibility is currently under study (Zhang et al., in preparation).

Again the relative contribution of two components might be responsible
for this trend. During the most intense events the flaring component
would completely dominate over the quasi--stationary one, progressively
shifting to higher energies (hard lag), while in weaker flares the
varying component would exceed the steady one only at energies higher
than the peak, where the spectrum steepens, producing an observed soft
lag behavior.

However, one should also consider that in different brightness states
the different observed energies correspond to different positions with
respect to E$_\mathrm{peak}$ (e.g. comparing the 1997 and 1998
datasets we showed that its value was typically a factor 2 different),
requiring a more subtle analysis before deriving inferences on the
lag--brightness relationship.


\subsection{Physical Interpretation: the signature of particle acceleration}
\label{sec:chiab}
\label{sec:interpretation}

Let us now focus on the interpretation of the two main and robust results
of this work, namely the hard lag and the evolution of the synchrotron peak.
 
The occurrence of the peak at different times for different energies is
most likely related to the particle acceleration/heating process. 
Although models to reproduce the temporal evolution of a spectral distribution 
have been developed (e.g. Chiaberge \& Ghisellini \citealp{chiab_gg}, 
Georganopoulos \& Marscher \citealp{markos_98}; 
Kirk, Rieger \& Mastichiadis \citealp{kirk_etal_98}),
they mostly do not consider the role of particle acceleration
(see however Kirk et al.~\citealp{kirk_etal_98}).

In order to account for the above results, we thus introduced a
(parametric) acceleration term in the particle kinetic equation within
of the time dependent model studied by Chiaberge \& Ghisellini
(\citealp{chiab_gg}).  In particular, this model takes into account the
cooling and escape in the evolution of the particle distribution and
the role of delays in the received photons due to the light crossing
time of different parts of the emitting region.

The model includes the presence of a quiescent spectrum, which is
assumed to be represented by and thus fitted to the broad band
spectral distribution observed in 1994 (Macomb et al.~\citealp{macomb95}).  
To this a flaring component is added and this is constrained by the
observed spectral and temporal evolution (the parameters for both the
stationary and variable spectra are reported in
Table~\ref{tab:model_parameters}).
The values of the model parameters are very close to those derived in 
Maraschi et al. (\citealp{maraschi_letter}) in a somewhat independent way,
i.e. trying only to fulfill the constraints provided by the 
X--ray and TeV spectra, simultaneous but averaged over the whole flare.
On this basis it is very likely that the model discussed here satisfies the
TeV constraints, which are however not directly taken into account
(the computation of the TeV light curve in this complex model would
require to treat the non--locality of the IC process, which is beyond the
scope of this paper).

Clearly a parametric prescription does not reproduce a priori a
specific acceleration process, but we rather tried to constrain its form
from the observed evolution.  The main constraints on the form of the
acceleration term are the following:

\begin{itemize}
\item[a)] Particles have to be injected at progressively higher
energies on the flare rise timescale to produce the hard lag. 

\item[b)] Globally the range of energies over which the injection occurs
has to be narrow, to give 
rise to a peaked spectral component.

\item[c)] A quasi--monochromatic injection function for the flux at
the highest energies to reach its maximum {\it after} that at lower
ones (as opposed to e.g. a power law with increasing maximum electron
Lorentz factor).

\item[d)] The emission in the LECS band from the particles which have
been accelerated to the highest energies (i.e. those radiating
initially in the MECS band) should not exceed that from the lower
energy ones, as after the peak no further increase of the (LECS) flux
is observed; this in particular requires for the injection to stop
after reaching the highest energies.

\item[e)] The total decay timescale might be dominated by the
achromatic crossing time effects, although the initial fading might be
determined by different cooling timescales.  
\end{itemize}

\begin{figure}[t]
\centerline{\includegraphics[width=0.90\linewidth]{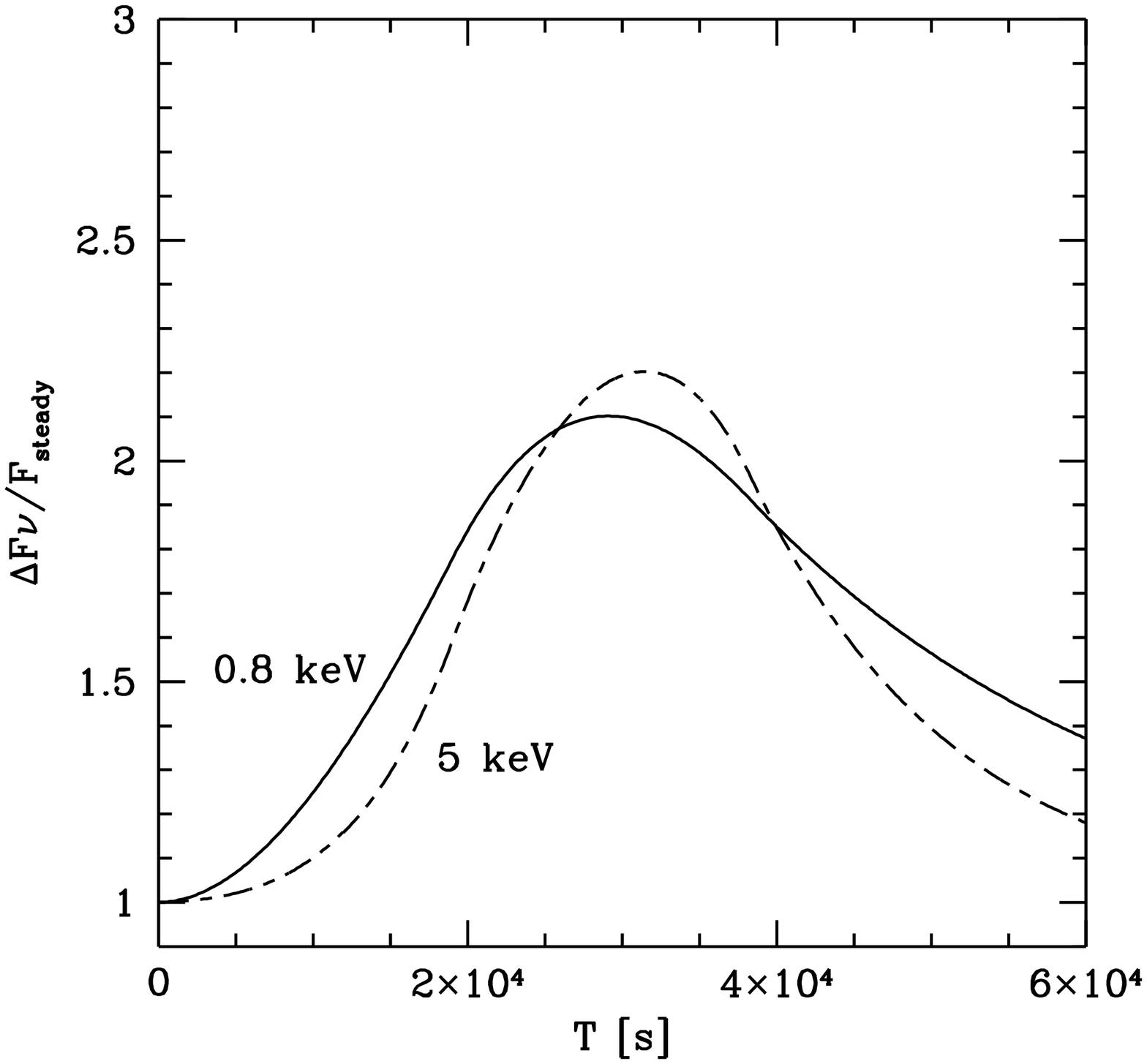}}
\vspace{-0.4cm}
\caption{\footnotesize\protect\baselineskip 8pt%
Light curves computed from the model, at the centroid energies of the soft and
hard band considered in the data analysis. The vertical axis
represents the variations normalized to the flux of the stationary
component.  \label{fig:model_lc} }
\centerline{\includegraphics[clip=,width=0.90\linewidth]{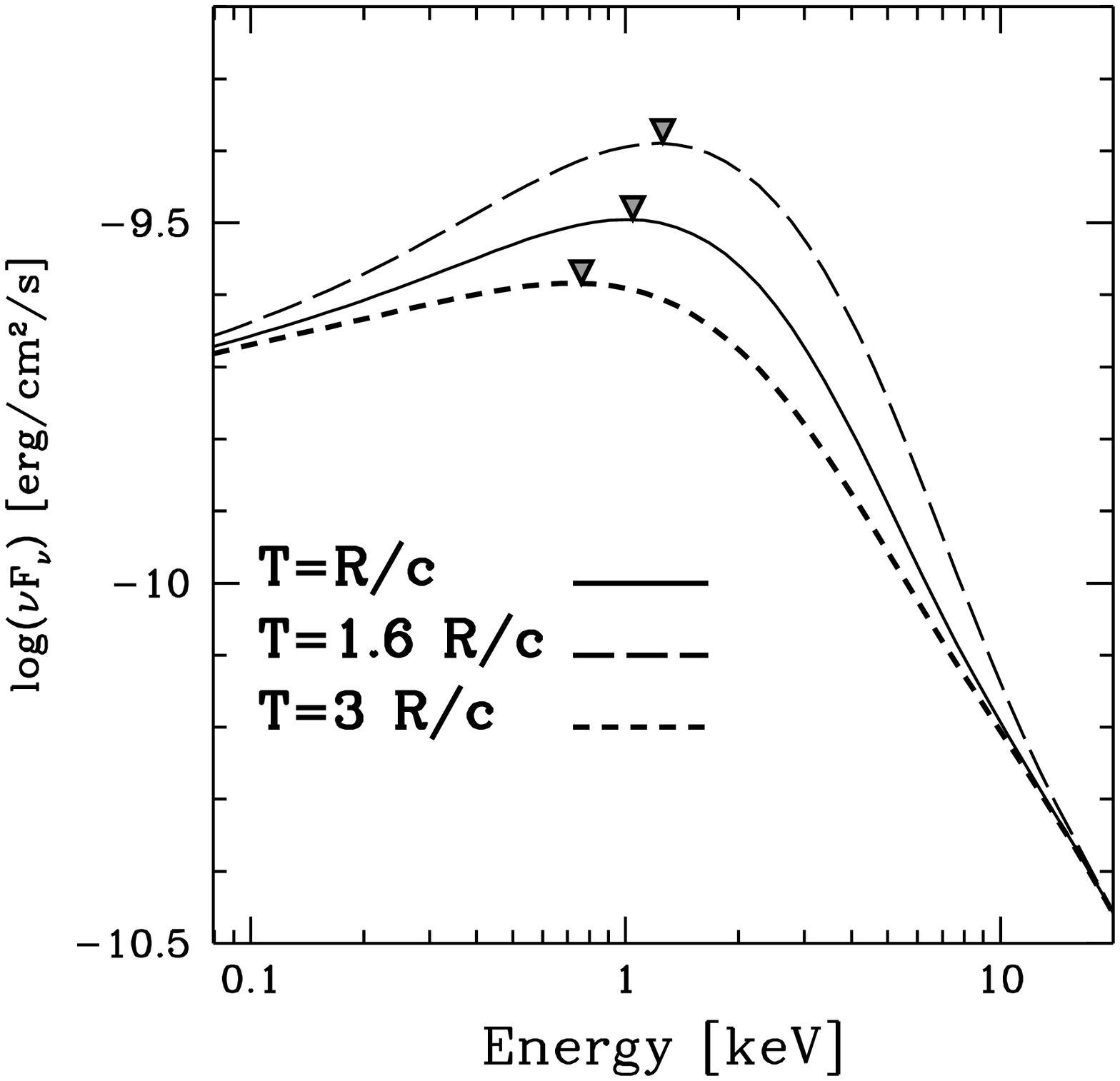}}
\vspace{-0.4cm}
\caption{\footnotesize\protect\baselineskip 8pt%
X--ray spectra at different times, as derived from the model. 
On each spectrum we marked the position of the synchrotron peak with a symbol.
\label{fig:model_spectra} }
\end{figure}

\input{Tab_Model_Parameters.txt}

\noindent
The acceleration term has then been described as a Gaussian distribution in
energy, centered at a typical particle Lorentz factor
$\gamma_\mathrm{c}(\mathrm{t})$ and with width $\sigma = 0.01
\gamma_\mathrm{c}(\mathrm{t})$, exponentially
evolving with time: $\gamma_\mathrm{c}(\mathrm{t}) \propto
e^{-(\mathrm{t}_\mathrm{max}-\mathrm{t})}$, where t$_\mathrm{max}$ corresponds
to the end of the acceleration phase, when the maximum
$\gamma_\mathrm{c,max}$ is reached. The injected luminosity is assumed to
be constant in time.

The duration of the injection, assumed to correspond to the light crossing
time of the emitting region, is such that it mimics the passage of a shock
front (i.e. a moving surface passing through the region in the same time
interval, see Chiaberge \& Ghisellini~\citealp{chiab_gg}).

It should be also noted that, within this scenario, the symmetry
between the rise and decay in the soft energy light curve seems to
suggest that if the energy where most of the power is released is
determined by the balance between the acceleration and cooling rates,
at this very same energy the latter timescales are comparable to the
light crossing time of the region.

The predictions of the model are shown in Figures~\ref{fig:model_lc}
and~\ref{fig:model_spectra}, in the form of light curves and spectra
at different times, respectively.  In particular, in
Figure~\ref{fig:model_lc} the light curves (normalized over the
stationary component) at the two centroid energies of the soft and
hard bands considered in the data analysis are reported.  The presence of a
hard lag is clearly visible as well as the larger variability
amplitude at higher frequencies.  Also the spectral evolution,
reported in Figure~\ref{fig:model_spectra} in $\nu F_\nu$ in the
observed energy range, seems to be at least in qualitative agreement
with what observed (compare with Fig.~\ref{fig:quattro_xseds}).

A posteriori it is important to notice that the inclusion of a
quasi--stationary component and of high energy emission (from
synchrotron self--Compton) are indeed crucial in order to reproduce
the retarded spectral variability at energies above a few keV.
This constitute a further difference with respect to the discussion of
Maraschi et al. (\citealp{maraschi_letter}), where a simplified description of
the time decay was used, based on evidence for an energy dependence
of the flare decay timescale (obtained modeling the decay 
with no baseline).

\section{Conclusions}
\label{sec:conclusions}

\textit{Beppo}SAX has observed Mkn~421 in 1997 and 1998.  We analyzed
and interpreted the combined spectral and temporal evolution in the
X--ray range.  During these observations the source has shown a large
variety of behaviors, both concerning the X--ray band itself and its
variability properties with respect to the $\gamma$--ray one,
providing us with a great wealth of information, but at the same time
revealing a richer than expected phenomenology.

Several important results follow from this work:

\begin{itemize}
\item[(a)] The X--ray and 2~TeV light curves peak simultaneously 
\textit{within one hour}, although the halving time in the
TeV band seems shorter than those at LECS and MECS
energies (see Maraschi et al. \citealp{maraschi_letter}).

\item[(b)] The detailed comparison of 0.1--1.5~keV and 3.5--10~keV
band light curves shows that the higher energy band lags the softer
one, with a delay of the order of 2--3~ks.  
This finding is opposite to what has been commonly observed in HBL X--ray
spectra (see Paper~I).

\item[(c)] Moreover, extracting LECS$+$MECS spectra for $\lesssim$
5~ks intervals, we were able to follow in detail the spectral
evolution during the flare. For the \textit{first} time we could 
quantitatively track the shift of the peak of the synchrotron component
moving to higher energy during the rising phase of the flare, and then
receding.

\item[(d)] An energy dependence of the shape of the light curve during
the flare has been revealed: at low energies the shape is consistent
with being symmetric, while at higher energies is clearly
asymmetric (faster rise) (see Paper~I).  

\item[(e)] Evidence has been found for the presence of the IC
component, and more importantly for its substantial variability, which
is possibly delayed with respect to the synchrotron one.
\end{itemize}

\noindent
These findings provide several temporal and spectral constraints on
any model. In particular, they seem to reveal the first direct
signature of the ongoing particle acceleration mechanism,
progressively ``pumping'' electrons from lower to higher energies.
The measure of the delay between the peaks of the light curves at the
corresponding emitted frequencies thus provides a tight constraint on the
timescale of the acceleration process.

Indeed, within a single emission region scenario, we have been able to
reproduce the sign and amount of lag by postulating that particle
acceleration follows a simple exponential law in time, stops at the
highest particle energies, and lasts for an interval comparable to the
light crossing time of the emitting region. If this timescale is
intrinsically linked to the typical source size, we indeed expect the
observed light curve to be symmetric at the energies where the bulk of
power is concentrated and an almost achromatic decay.  The same model
can account for the spectral evolution (shift of the synchrotron peak)
during the flare.

The other very important clue derived from the analysis is the
presence of a quasi--stationary contribution to the emission, which seems to
be dominated by a highly variable peaked spectrum, possibly
maintaining a quasi--rigid shape during flares. The decomposition of
the observed spectrum into these two components might allow us to
determine the nature and modality of the energy dissipation in
relativistic jets.

\acknowledgments

We are grateful to the \textit{Beppo}SAX Science Data Center (SDC) for their 
invaluable work and for providing standardized product data archive, 
and to the \textit{Rossi}XTE ASM Team.
We thank Gianpiero Tagliaferri and Paola Grandi for their contribution to
our successful \textit{Beppo}SAX program, and for useful comments
and the anonymous referee for suggestions which have
improved the clarity of the paper.
AC, MC and YHZ acknowledge the Italian MURST for financial support. 
This research was supported in part by the National Science Foundation
under Grant No.~PHY94--07194 (AC).
Finally, GF thanks Cecilia Clementi for providing tireless stimulus.

\appendix 
\section{A. A caveat on continuously curved spectral models}
\label{sec:appendix:curved_model}

The increasing quality of the available X--ray spectral data, both in
terms of signal--to--noise ratio and energy resolution, has made
necessary to consider more complex fitting models. One of the most
interesting continuum feature in the X--ray spectra of blazars is the
curvature of the synchrotron component, as good quality data could
enable us to estimate the energy of the emission peak.  As clearly the
traditional single or broken power law models are unsatisfactory, we
developed the curved spectral model presented in this paper: a simple
analytic expression representing a continuous curvature, two
``pivoting'' points --useful for analysis purposes-- asymptotically
joining two power law branches, whose slope is completely determined
by the behavior of the function at the pivoting energies.

Another increasingly popular model has been introduced by Giommi et
al.~(\citealp{giommi_2155}) (to reproduce \textit{Beppo}SAX data of
PKS~2155$-$304, and then used also for Mkn~421 by Guainazzi et
al.~\citealp{guainazzi_mkn421} and Malizia et al.~\citealp{malizia_mkn421}),
but unfortunately this is affected by a problem that can lead to
misleading results.

This more generally occurs for all curved continuum models defined as:
\begin{equation}
F(E) \; = \; E^{-\tilde\alpha(E)}
\end{equation}
where the function $\alpha(E)$ is suitably defined to smoothly
change between two asymptotic values $\alpha_1$ and $\alpha_2$.
The specific choice by Giommi et al.~(\citealp{giommi_2155}) is:
\begin{equation}
\tilde\alpha(E) \; = \; f(E) \alpha_1 + (1 - f(E)) \alpha_2
\end{equation}
with
\begin{equation}
f(E) \; = \; \left(1 - e^{-E/E_0}\right)^\beta.
\end{equation}
Thus $\tilde\alpha \rightarrow \alpha_1$ for $E \ll E_0$, and
$\tilde\alpha \rightarrow \alpha_2$ for $E \gg E_0$.  While this
exactly the behavior of the spectral index function, this does not
correspond to the real spectral index of the function $F(E)$,
conventionally defined as:
\begin{equation}
\alpha(E) \;\equiv\; - \frac{d \ln F(E)}{d \ln E}.
\end{equation}
In fact for $F(E) \equiv E^{-\tilde\alpha(E)}$ this yields
\begin{equation}
\alpha(E) \;=\; \tilde\alpha(E) \;+\; \ln E\, \frac{d \tilde\alpha(E)}{d \ln E}  
          \;=\; \tilde\alpha(E) \;+\; \delta_{\tilde\alpha}(E)
\end{equation}
thus including an unavoidable term $\delta_{\tilde\alpha}(E)$, 
present because of the not vanishing (by definition) derivative of
$\tilde\alpha(E)$.

For the above specific choice of $\tilde\alpha(E)$:
\begin{equation}
\delta_{\tilde\alpha}(E) \;=\; \beta\; \frac{E}{E_0} \ln E\; (\alpha_1 - \alpha_2)\; e^{-E/E_0} \left(1 -
e^{-E/E_0}\right)^{\beta-1}.
\end{equation}
The real spectral index then takes a value (sensibly) different
from the one expected over a large range of energies.

\begin{figure}[t]
\centerline{\includegraphics[width=0.90\linewidth]{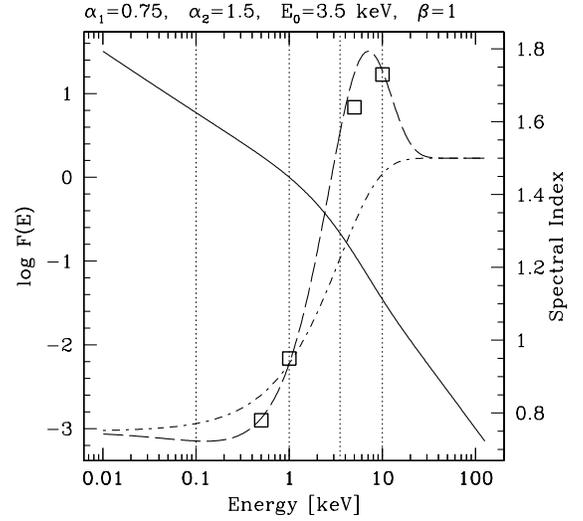}}
\vspace{-0.6cm}
\caption{\footnotesize\protect\baselineskip 8pt%
Example of the fitting spectral model used by Giommi et al.~(\citealp{giommi_2155}), for a choice
of model parameters suited to reproduce the spectral properties of the 1998
flare (shown on top of the figure box).
The solid line is the spectrum (scale on the left Y--axis). 
The dot--dashed line is the predicted  spectral index
$\tilde\alpha(E)$, and the dashed line is the ``true'' spectral index of
the resulting $F(E)$ (both with the scale on the right Y--axis).
The squares represent the values of the spectral index at the same energies
reported in Table~\ref{tab:spectral_analysis_98}.
Vertical bars indicate a few relevant reference energies, namely
0.1 and 10~keV, that are the limits of the band covered by
LECS and MECSs, 1~keV as the energy at which the model
is normalized to unity, and $E_0$.
}
\label{fig:esempio_model_matt}
\end{figure}

The effect can be better illustrated with an example.
Let us consider a set of spectral parameters that would well reproduce the
spectrum of the peak of the 1998 flare of Mkn~421: $\alpha_1 = 0.75,
\alpha_2 = 1.5, E_0 = 3.5$~keV, $\beta = 1.0$.
In Figure~\ref{fig:esempio_model_matt} we show the spectrum together with
$\tilde\alpha(E)$ and $\alpha(E)$ as derived by the model over an energy interval
broader than that effectively covered by data.
For reference we also plotted the values of $\alpha$ derived from our curved
model fits, to show that the real spectral index $\alpha(E)$ matches
them.  A few features emerge:
\begin{enumerate}
\item according to $\tilde\alpha(E)$
between $E_0$ and 10~keV there is an apparent  wiggle in the spectrum, which
steepens beyond the supposedly asymptotic value $\alpha_2$ before reaching
it at about 30~keV.
\item within the 0.1--10~keV range $\alpha(E)$ and $\tilde\alpha(E)$ never
match each other except for $E = 1$~keV, when $\delta_{\tilde\alpha}(E)$
vanishes.
\item the range spanned by the true spectral index is broader, and
in particular it is always steeper than the value expected on the basis of
$\tilde\alpha(E)$ in the whole band 1--20~keV.
$\Delta\alpha$ can be as much as $+$0.42 (at $E \simeq 5.5$~keV).
\item at no energy in the 0.1--10~keV band the actual spectrum has a slope equal
to either $\alpha_1$ or $\alpha_2$, and thus these two parameters 
do not have any real descriptive meaning (this is strikingly illustrated
by the displacement of the real data points from the expected curve).
\item the total change of slope in the 0.1--10~keV interval would
be expected $\Delta\tilde\alpha \simeq 0.68$ while it actually turns out 
to be 1.04 (and even more dramatically between 0.5 and 5.0~keV,
$\Delta\tilde\alpha \simeq 0.46$ while for the data it is 0.86; see
Table~\ref{tab:spectral_analysis_98}).
\end{enumerate}
Summarizing, there is no way to define a curved continuum model by
means of a suitable function describing the power law exponent.

\section{B. More details on the adopted curved model}
\label{sec:appendix:this_curved_model}

The relationship between $\alpha(E)$ and $\alpha_{-\infty}$ and
$\alpha_{+\infty}$ is:
\begin{equation}
\alpha(E) \;\equiv\; - \frac{d \ln F(E)}{d \ln (E)} \;=\; 
\alpha_{-\infty} - (\alpha_{-\infty} - \alpha_{+\infty})\;
\frac{\left(E/E_\mathrm{B}\right)^f}{1 + \left(E/E_\mathrm{B}\right)^f}
\end{equation}
\noindent
yielding:
\begin{eqnarray}
\alpha_{-\infty} & \;=\; &
\frac{B\; (1+A)\; \alpha_1 - A\; (1+B)\; \alpha_2}{B-A} \\
\alpha_{+\infty} & \;=\; &
\frac{(1+B)\; \alpha_2 - (1+A)\; \alpha_1}{B-A} 
\end{eqnarray}
\noindent
where $A \;=\; \left(E_1/E_\mathrm{B}\right)^f$, and 
$B \;=\; \left(E_2/E_\mathrm{B}\right)^f$.

\section{C. More details on the broken power law fits.}
\label{sec:appendix:bkn_pl_fits}
\input{Tab_bknPL_Fits.txt}
\clearpage


\end{document}

%% file: Tab_Curved_Fits.txt
\begin{deluxetable}{cccccccccccc}
\tablecolumns{12}
\tablewidth{0pc}
\tablecaption{LECS and MECS Curved Model Spectral Fit: 1997 and 1998 \label{tab:spectral_analysis_97} \label{tab:spectral_analysis_98}}
\tablehead{
\colhead{Orbits}    &  
\colhead{Median Obs. Time} &
\multicolumn{4}{c}{spectral indices} &   
\colhead{$E_{\rm B}$}  & 
\colhead{$E_{\rm peak}$}  &  
\multicolumn{3}{c}{fluxes\tablenotemark{b}} &
\colhead{$\chi^2_\nu$\tablenotemark{c}}  \\
\cline{3-6} \cline{9-11} 
\colhead{}  &  
\colhead{UTC} &
\colhead{0.5 keV}  &  
\colhead{1 keV}  &  
\colhead{5 keV}  &  
\colhead{10 keV}  &  
\colhead{(keV)} &
\colhead{(keV)} &
\colhead{0.2-1} &
\colhead{2-10} &
\colhead{0.1-10} &
\colhead{} }
\startdata
\multicolumn{12}{c}{1997 April 29$^\mathrm{th}$ to May 5$^\mathrm{th}$} \\
\tableline
1--3 & 1997/04/29:06:08 & 0.91$^{+0.04}_{-0.04}$ & 1.19$^{+0.05}_{-0.04}$ & 1.83$^{+0.05}_{-0.05}$ & 1.99$^{+0.13}_{-0.10}$ & 1.27$^{+1.15}_{-0.62}$ & 0.62$^{+0.05}_{-0.05}$ & 2.09 & 1.01 & \phn4.68 & 1.24  \\ \smallskip
4--7 & 1997/04/29:11:43 & 1.00$^{+0.03}_{-0.03}$ & 1.28$^{+0.04}_{-0.03}$ & 1.93$^{+0.05}_{-0.05}$ & 2.09$^{+0.11}_{-0.09}$ & 1.34$^{+0.96}_{-0.55}$ & 0.49$^{+0.04}_{-0.05}$ & 1.89 & 0.75 & \phn4.07 & 1.03  \\ \smallskip
8--10 & 1997/04/30:05:01 & 0.94$^{+0.05}_{-0.06}$ & 1.24$^{+0.08}_{-0.06}$ & 1.91$^{+0.07}_{-0.06}$ & 2.08$^{+0.20}_{-0.13}$ & 1.26$^{+1.95}_{-0.79}$ & 0.57$^{+0.07}_{-0.08}$ & 2.15 & 0.92 & \phn4.65 & 0.95  \\ \smallskip
11--12 & 1997/04/30:08:37 & 1.01$^{+0.04}_{-0.04}$ & 1.24$^{+0.05}_{-0.04}$ & 1.91$^{+0.07}_{-0.06}$ & 2.11$^{+0.21}_{-0.14}$ & 1.99$^{+2.65}_{-1.08}$ & 0.48$^{+0.07}_{-0.11}$ & 1.96 & 0.83 & \phn4.29 & 0.94  \\ \smallskip
13--15 & 1997/04/30:12:36 & 0.95$^{+0.03}_{-0.03}$ & 1.26$^{+0.04}_{-0.04}$ & 1.95$^{+0.05}_{-0.05}$ & 2.13$^{+0.12}_{-0.09}$ & 1.25$^{+0.90}_{-0.53}$ & 0.55$^{+0.04}_{-0.04}$ & 1.92 & 0.78 & \phn4.14 & 0.91  \\ \smallskip
16--18 & 1997/05/01:05:08 & 1.03$^{+0.05}_{-0.05}$ & 1.35$^{+0.07}_{-0.07}$ & 1.84$^{+0.07}_{-0.06}$ & 1.94$^{+0.13}_{-0.09}$ & 0.57$^{+0.91}_{-0.42}$ & 0.46$^{+0.05}_{-0.07}$ & 1.99 & 0.76 & \phn4.20 & 1.14  \\ \smallskip
19--20 & 1997/05/01:08:43 & 1.00$^{+0.04}_{-0.04}$ & 1.25$^{+0.05}_{-0.05}$ & 1.78$^{+0.06}_{-0.05}$ & 1.91$^{+0.14}_{-0.10}$ & 1.14$^{+1.51}_{-0.66}$ & 0.49$^{+0.06}_{-0.08}$ & 2.00 & 0.89 & \phn4.42 & 1.31  \\ \smallskip
21--23 & 1997/05/01:12:39 & 0.92$^{+0.03}_{-0.03}$ & 1.20$^{+0.04}_{-0.03}$ & 1.86$^{+0.04}_{-0.04}$ & 2.04$^{+0.11}_{-0.08}$ & 1.44$^{+0.98}_{-0.58}$ & 0.61$^{+0.04}_{-0.04}$ & 2.60 & 1.22 & \phn5.79 & 1.21  \\ \smallskip
24--25 & 1997/05/02:05:36 & 0.93$^{+0.04}_{-0.04}$ & 1.18$^{+0.05}_{-0.04}$ & 1.83$^{+0.06}_{-0.05}$ & 2.00$^{+0.15}_{-0.11}$ & 1.57$^{+1.64}_{-0.80}$ & 0.61$^{+0.06}_{-0.07}$ & 2.74 & 1.34 & \phn6.20 & 0.77  \\ \smallskip
26--27 & 1997/05/02:08:35 & 0.93$^{+0.05}_{-0.05}$ & 1.24$^{+0.06}_{-0.05}$ & 1.88$^{+0.06}_{-0.06}$ & 2.03$^{+0.15}_{-0.10}$ & 1.08$^{+1.27}_{-0.60}$ & 0.58$^{+0.06}_{-0.06}$ & 2.56 & 1.12 & \phn6.20 & 1.03  \\ \smallskip
28--30 & 1997/05/03:05:21 & 1.07$^{+0.05}_{-0.05}$ & 1.30$^{+0.07}_{-0.05}$ & 1.88$^{+0.07}_{-0.07}$ & 2.05$^{+0.22}_{-0.14}$ & 1.63$^{+3.04}_{-1.05}$ & 0.37$^{+0.09}_{-0.19}$ & 2.01 & 0.78 & \phn4.35 & 1.13  \\ \smallskip
31--32 & 1997/05/03:08:38 & 1.05$^{+0.06}_{-0.06}$ & 1.28$^{+0.06}_{-0.05}$ & 2.13$^{+0.10}_{-0.09}$ & 2.49$^{+0.39}_{-0.24}$ & 3.50$^{+6.97}_{-2.05}$ & 0.39$^{+0.12}_{-0.19}$ & 1.65 & 0.62 & \phn3.63 & 0.79  \\ \smallskip
33--34 & 1997/05/04:04:22 & 1.11$^{+0.07}_{-0.07}$ & 1.47$^{+0.13}_{-0.10}$ & 2.15$^{+0.15}_{-0.12}$ & 2.30$^{+0.37}_{-0.20}$ & 0.91$^{+2.84}_{-0.75}$ & 0.39$^{+0.07}_{-0.20}$ & 1.59 & 0.42 & \phn3.16 & 1.27  \\ \smallskip
35--37 & 1997/05/04:07:55 & 1.21$^{+0.05}_{-0.05}$ & 1.49$^{+0.08}_{-0.06}$ & 2.14$^{+0.10}_{-0.08}$ & 2.31$^{+0.25}_{-0.16}$ & 1.38$^{+2.42}_{-0.88}$ & 0.23$^{+0.11}_{-0.22}$ & 1.60 & 0.40 & \phn3.25 & 0.92  \\ \smallskip
38--40 & 1997/05/05:05:22 & 1.10$^{+0.04}_{-0.04}$ & 1.43$^{+0.07}_{-0.06}$ & 2.01$^{+0.07}_{-0.07}$ & 2.13$^{+0.15}_{-0.11}$ & 0.74$^{+1.01}_{-0.47}$ & 0.39$^{+0.05}_{-0.09}$ & 2.04 & 0.62 & \phn4.16 & 1.07  \\ \smallskip
41--42 & 1997/05/05:08:46 & 1.10$^{+0.06}_{-0.06}$ & 1.35$^{+0.08}_{-0.06}$ & 1.98$^{+0.10}_{-0.08}$ & 2.15$^{+0.29}_{-0.16}$ & 1.61$^{+3.73}_{-1.08}$ & 0.33$^{+0.10}_{-0.24}$ & 1.88 & 0.65 & \phn4.00 & 0.62  \\ 
\cutinhead{1998 April 21$^\mathrm{st}$}
1--2 & 1998/04/21:02:55 & 0.83$^{+0.05}_{-0.06}$ & 1.10$^{+0.07}_{-0.06}$ & 1.64$^{+0.06}_{-0.06}$ & 1.68$^{+0.09}_{-0.07}$ & 1.19$^{+0.47}_{-0.34}$ & 0.80$^{+0.10}_{-0.08}$ & 5.10 & 3.00 & 11.96 &  0.91   \\ \smallskip
3    & 1998/04/21:05:07 & 0.87$^{+0.05}_{-0.05}$ & 1.02$^{+0.04}_{-0.04}$ & 1.65$^{+0.08}_{-0.07}$ & 1.74$^{+0.15}_{-0.10}$ & 1.82$^{+0.74}_{-0.49}$ & 0.92$^{+0.12}_{-0.10}$ & 5.34 & 3.46 & 13.08 &  0.97   \\ \smallskip
4    & 198/04/21:06:40 & 0.78$^{+0.04}_{-0.04}$ & 0.95$^{+0.03}_{-0.03}$ & 1.64$^{+0.06}_{-0.06}$ & 1.73$^{+0.11}_{-0.09}$ & 1.85$^{+0.51}_{-0.38}$ & 1.13$^{+0.11}_{-0.10}$ & 5.96 & 4.41 & 15.22 &  1.02   \\ \smallskip
5    & 1998/04/21:08:19 & 0.81$^{+0.04}_{-0.04}$ & 0.96$^{+0.03}_{-0.03}$ & 1.58$^{+0.06}_{-0.05}$ & 1.66$^{+0.11}_{-0.08}$ & 1.82$^{+0.64}_{-0.44}$ & 1.09$^{+0.12}_{-0.10}$ & 5.77 & 4.22 & 14.63 &  0.86   \\ \smallskip
6    & 1998/04/21:09:56 & 0.80$^{+0.04}_{-0.04}$ & 1.03$^{+0.05}_{-0.04}$ & 1.50$^{+0.05}_{-0.05}$ & 1.53$^{+0.07}_{-0.06}$ & 1.16$^{+0.40}_{-0.30}$ & 0.91$^{+0.10}_{-0.09}$ & 5.35 & 3.76 & 13.24 &  1.17   \\ \smallskip
7    & 1998/04/21:11:32 & 0.79$^{+0.04}_{-0.04}$ & 1.00$^{+0.05}_{-0.04}$ & 1.53$^{+0.06}_{-0.05}$ & 1.57$^{+0.08}_{-0.07}$ & 1.36$^{+0.46}_{-0.33}$ & 0.98$^{+0.11}_{-0.10}$ & 5.05 & 3.62 & 12.64 &  0.85   \\ \smallskip
8--9 & 1998/04/21:13:57 & 0.85$^{+0.03}_{-0.03}$ & 1.05$^{+0.04}_{-0.03}$ & 1.50$^{+0.04}_{-0.04}$ & 1.54$^{+0.06}_{-0.05}$ & 1.26$^{+0.35}_{-0.27}$ & 0.85$^{+0.07}_{-0.06}$ & 4.62 & 3.10 & 11.35 &  1.38   \\ \smallskip
10--12 & 1998/04/21:17:57 & 0.87$^{+0.02}_{-0.02}$ & 1.05$^{+0.03}_{-0.02}$ & 1.54$^{+0.04}_{-0.03}$ & 1.58$^{+0.05}_{-0.04}$ & 1.40$^{+0.30}_{-0.24}$ & 0.85$^{+0.06}_{-0.05}$ & 3.99 & 2.63 & \phn9.76 &  0.89   \\ \smallskip
13--16 & 1998/04/21:23:35 & 0.91$^{+0.03}_{-0.03}$ & 1.06$^{+0.03}_{-0.02}$ & 1.61$^{+0.05}_{-0.05}$ & 1.67$^{+0.08}_{-0.07}$ & 1.67$^{+0.48}_{-0.36}$ & 0.79$^{+0.07}_{-0.07}$ & 3.61 & 2.23 & \phn8.73 &  0.99   \\ 
\cutinhead{1998 April 23$^\mathrm{rd}$}
\smallskip
1--5 & 1998/04/23:03:49 & 0.93$^{+0.02}_{-0.03}$ & 1.10$^{+0.03}_{-0.02}$ & 1.67$^{+0.04}_{-0.04}$ & 1.74$^{+0.07}_{-0.05}$ & 1.59$^{+0.37}_{-0.29}$ & 0.71$^{+0.05}_{-0.05}$ & 4.15 & 2.32 & \phn9.74 &  1.11   \\ \smallskip
6--11 & 1998/04/23:12:31 & 0.95$^{+0.02}_{-0.02}$ & 1.17$^{+0.02}_{-0.02}$ & 1.78$^{+0.03}_{-0.03}$ & 1.84$^{+0.05}_{-0.04}$ & 1.47$^{+0.21}_{-0.18}$ & 0.60$^{+0.03}_{-0.04}$ & 3.73 & 1.78 & \phn8.42 &  1.27   \\ \smallskip
12--19 & 1998/04/23:23:39 & 1.03$^{+0.02}_{-0.02}$ & 1.19$^{+0.02}_{-0.02}$ & 1.79$^{+0.04}_{-0.04}$ & 1.85$^{+0.06}_{-0.05}$ & 1.68$^{+0.32}_{-0.26}$ & 0.38$^{+0.09}_{-0.23}$ & 3.25 & 1.47 & \phn7.30 &  1.16   \\ 
\enddata
\tablenotetext{a}{\footnotesize 
Since there are always three varying parameters --either two $\alpha$ and
$E_\mathrm{B}$ or one $\alpha$, one $E$ and $E_\mathrm{B}$-- all the quoted
errors are for 1 $\sigma$ for 3 parameters (i.e. $\Delta\chi^2 = 3.53$), a
quite conservative choice.}
\tablenotetext{b}{\footnotesize Unabsorbed Fluxes 
in units of 10$^{-10}$ erg cm$^{-2}$ s$^{-1}$, in the reported
energy band, whose boundaries are expressed in keV.}
\tablenotetext{c}{\footnotesize The {\it degrees of freedom} are 57 for 
all cases} \end{deluxetable}

%% file: Tab_PL_Fits.txt
\begin{deluxetable}{ccccccccccc}
\tablecolumns{11}
\tablewidth{0pc}
\tablecaption{Summary of Power Law models Results \label{tab:power_laws_fit}}
\tablehead{
\colhead{}  &  
\colhead{}    &  
\multicolumn{4}{c}{1997} &   
\colhead{}    &
\multicolumn{4}{c}{1998} \\
\cline{3-6} \cline{8-11} 
\colhead{Model}  &  
\colhead{}  &  
\colhead{N$_\mathrm{H}$\tablenotemark{a}} &
\colhead{$\chi^2$}  &  
\colhead{$\chi^2_\nu$}  &  
\colhead{\it d.o.f.}  &  
\colhead{}  &  
\colhead{N$_\mathrm{H}$\tablenotemark{a}} &
\colhead{$\chi^2$}  &  
\colhead{$\chi^2_\nu$} &
\colhead{\it d.o.f.} }
\startdata
single P.L. && 
Galactic & 11903.1 &    12.609 & 944 && 
Galactic & 11245.0 &    15.883 & 708 \\ 
single P.L. && 
$3.78 \pm 0.18$ & \phn3012.7 & \phn3.246 & 928 && 
$3.22 \pm 0.14$ & \phn3413.2 & \phn4.905 & 696 \\ 
broken P.L. && 
Galactic & \phn1162.6 & \phn1.275 & 912 && 
Galactic & \phn\phn997.6 & \phn1.458 & 684 \\ 
broken P.L. &&
$2.23 \pm 0.46$ & \phn1083.2 & \phn1.209 & 896 && 
$2.05 \pm 0.19$ & \phn\phn866.6 & \phn1.289 & 672 \\ 
curved      && 
Galactic & \phn\phn935.7 & \phn1.026 & 912 && 
Galactic & \phn\phn721.1 & \phn1.054 & 684 \\  
curved      && 
$1.73 \pm 0.12$ & \phn\phn917.8 & \phn1.024 & 896 && 
$1.70 \pm 0.12$ & \phn\phn705.2 & \phn1.049 & 672 \\ 
\enddata
\tablenotetext{a}{\footnotesize absorbing column units are $10^{20}$ cm$^{-2}$.}
\end{deluxetable}

%% file: Tab_PDS_fits1.txt
\begin{deluxetable}{lcccccccc}
\tablecolumns{9}
\tablewidth{0pc}
\tablecaption{PDS single power law fits: 1997 and 1998 \label{tab:pds_spectra}}
\tablehead{
\colhead{} & 
\colhead{} & 
\colhead{} &
\multicolumn{2}{c}{Count rates} &  
\colhead{} & 
\colhead{} & 
\colhead{} & 
\colhead{} \\
\cline{4-5} 
\colhead{Dataset}    &  
\colhead{Band} & 
\colhead{Exposure}   &  
\colhead{12--40 keV}    &
\colhead{40--90 keV}    &
\colhead{$\alpha$\tablenotemark{a}}   &  
\colhead{F$_\mathrm{1 keV}$}   &  
\colhead{$\chi^2$} &
\colhead{\textit{d.o.f.}} \\
\colhead{} &
\colhead{} &
\colhead{(ks)} &
\colhead{(cts/s)} &
\colhead{(cts/s)} &
\colhead{} &
\colhead{($\mu$Jy)} &
\colhead{} &
\colhead{} }
\startdata
1997                        & 12--90~keV & 54.8 & 0.177 $\pm$ 0.021 & 0.045 $\pm$ 0.019 & 2.17$^{+0.48}_{-0.41}$ & 1.52 & 2.96 & 3 \\[0.1cm]
1998 April 21$^\mathrm{st}$ & 12--90~keV & 21.1 & 0.698 $\pm$ 0.028 & 0.039 $\pm$ 0.023 & 2.23$^{+0.14}_{-0.13}$ & 5.76 & 1.76 & 3 \\[0.1cm]
1998 April 23$^\mathrm{rd}$ & 12--90~keV & 25.2 & 0.293 $\pm$ 0.025 & 0.081 $\pm$ 0.021 & 1.35$^{+0.26}_{-0.24}$ & 1.35 & 6.95 & 4 \\[0.1cm]
                            & 12--130~keV & \nodata & \nodata           & \nodata           & 1.54$^{+0.28}_{-0.25}$ & 1.54 & 2.95 & 3 \\[0.1cm]
\enddata
\tablenotetext{a}{\footnotesize 
Errors are for 1 $\sigma$ for 1 parameter (i.e. $\Delta\chi^2 = 1.0$).}
\end{deluxetable}

%% file: Tab_PDS_fits2.txt
\begin{deluxetable}{lccccccc}
\tablecolumns{8}
\tablewidth{0pc}
\tablecaption{High energy component constraints from MECS$+$PDS spectral fits \label{tab:ic_pl_fits}}
\tablehead{
\colhead{}    &  
\multicolumn{2}{c}{High Energy Component Flux\tablenotemark{a,b}} &  
 &
\multicolumn{2}{c}{Synchrotron Flux\tablenotemark{a,c}} &  
\colhead{} &
\colhead{}\\
\cline{2-3} \cline{5-6}
\colhead{Dataset}    &  
\colhead{12--40 keV} & \colhead{40--100 keV} &
 &
\colhead{12--40 keV} & \colhead{40--100 keV} &
\colhead{$\chi^2$} &
\colhead{\textit{d.o.f.}} }
\startdata
1997                         & 
\phantom{$<$} 11.0$^{+2.1}_{-2.6}$ & \phantom{$<$} 12.8$^{+3.7}_{-2.1}$ &&  \phn7.8 & 0.2 & 17.92 & 15 \\[0.4cm]
1998 April 21$^\mathrm{st}$  & 
$<$ 13.0\tablenotemark{d}\phantom{$^{+2.1}_{-2.6}$} & $<$ 14.9\tablenotemark{d}\phantom{$^{+2.1}_{-2.6}$} && 78.4 & 8.6 & 28.71 & 30 \\[0.1cm]
1998 April 23$^\mathrm{rd}$  &
\phantom{$<$} 13.6$^{+3.3}_{-3.1}$ & \phantom{$<$} 15.8$^{+3.8}_{-3.6}$ && 24.0 & 1.2 & \nodata & \nodata \\[0.4cm]
1998 April 21$^\mathrm{st}$ $+$ 23$^\mathrm{rd}$ & 
\phantom{$<$} 10.1$^{+3.1}_{-2.9}$ & \phantom{$<$} 11.8$^{+3.5}_{-3.4}$ && 78.4~/~24.0 & 8.6~/~1.2 & 34.79 & 31 \\[0.1cm]
\enddata
\tablenotetext{a}{\footnotesize 
Fluxes are in units of 10$^{-12}$ erg cm$^{-2}$ s$^{-1}$.
Errors are for 1 $\sigma$ for 1 parameter (i.e. $\Delta\chi^2 = 1.0$).}
\tablenotetext{b}{\footnotesize 
Flux computed by the best fit model in the hard power law. }
\tablenotetext{c}{\footnotesize 
Flux computed by the best fit model in the curved model component. }
\tablenotetext{d}{\footnotesize 
90\% upper limit. }
\end{deluxetable}

%% file: Tab_Model_Parameters.txt
\begin{deluxetable}{clcc}
\tablecolumns{4}
\tablewidth{0pc}
\tablecaption{Model Parameters\label{tab:model_parameters}}
\tablehead{
\colhead{Parameter} & \colhead{Parameter Description} & \colhead{stationary} & \colhead{flaring} 
}
\startdata
$R$                    & source dimension (cm)  & 2 $\cdot 10^{16}$    & 2 $\cdot 10^{16}$   \\
$B$                    & magnetic field (G)     & 0.12     & 0.03  \\
$t_{esc}$              & escape timescale (R/c) & 6        & 1.5   \\
$\ell_\mathrm{inj}$    & compactness injected as particle energy & 6 $\cdot 10^{-5}$     & 3 $\cdot 10^{-4}$  \\
$s$                    & slope of the injected stationary power law & 1.6      & \nodata \\
$\gamma_\mathrm{max}$  & maximum Lorentz factor of the injected stationary power law & 3 $\cdot 10^5$   & \nodata \\
$\gamma_\mathrm{c,max}$& maximum Lorentz factor of Gaussian injection for the flaring component & \nodata  & 7 $\cdot 10^5$ \\
$\delta$               & beaming factor & 18       & 18 \\
\enddata
\end{deluxetable}

%% file: Tab_bknPL_Fits.txt
\begin{deluxetable}{cccccccccccc}
\tablecolumns{11}
\tablewidth{0pc}
\tablecaption{LECS and MECS Broken Power Law Model Spectral Fit: 1997 and 1998\label{tab:spectral_analysis_bkn_97} \label{tab:spectral_analysis_bkn_98}}
\tablehead{
\colhead{Orbits}    &  
\multicolumn{4}{c}{Galactic absorbing column\tablenotemark{a}} &   
\colhead{}    &  
\multicolumn{5}{c}{free absorbing column\tablenotemark{b}} \\
\cline{2-5} \cline{7-11} \\
\colhead{}    &  
\colhead{$\alpha_1$}    &  
\colhead{$\alpha_2$}    &  
\colhead{E$_\mathrm{break}$}  & 
\colhead{$\chi^2_\nu$}    &  
\colhead{}    &  
\colhead{N$_\mathrm{H}$}    &  
\colhead{$\alpha_1$}    &  
\colhead{$\alpha_2$}    &  
\colhead{E$_\mathrm{break}$}  & 
\colhead{$\chi^2_\nu$} \\
\colhead{}    &  
\colhead{}    &  
\colhead{}    &  
\colhead{(keV)} &
\colhead{}    &  
\colhead{}    &  
\colhead{(10$^{20}$ cm$^{-2}$)}    &  
\colhead{}    &  
\colhead{}    &  
\colhead{(keV)} &
\colhead{} }
\startdata
\multicolumn{11}{c}{1997 April 29$^\mathrm{th}$ to May 5$^\mathrm{th}$} \\
\tableline
   1--3 & 0.96$^{+0.05}_{-0.07}$ & 1.71$^{+0.04}_{-0.03}$ & 1.37$^{+0.18}_{-0.19}$ & 1.71&& 2.50$^{+0.32}_{-0.30}$ & 1.26$^{+0.08}_{-0.08}$ & 1.77$^{+0.06}_{-0.05}$ & 2.25$^{+0.23}_{-0.23}$ & 1.37 \\ \smallskip 
   4--7 & 1.01$^{+0.04}_{-0.04}$ & 1.79$^{+0.03}_{-0.03}$ & 1.23$^{+0.12}_{-0.10}$ & 1.55&& 1.96$^{+0.37}_{-0.41}$ & 1.14$^{+0.13}_{-0.17}$ & 1.80$^{+0.04}_{-0.04}$ & 1.38$^{+0.27}_{-0.21}$ & 1.51 \\ \smallskip 
  8--10 & 0.93$^{+0.07}_{-0.09}$ & 1.77$^{+0.04}_{-0.04}$ & 1.19$^{+0.17}_{-0.15}$ & 1.38&& 2.92$^{+0.43}_{-0.43}$ & 1.40$^{+0.10}_{-0.10}$ & 1.95$^{+0.13}_{-0.11}$ & 2.90$^{+0.48}_{-0.51}$ & 1.23 \\ \smallskip 
 11--12 & 1.04$^{+0.05}_{-0.05}$ & 1.77$^{+0.04}_{-0.04}$ & 1.41$^{+0.18}_{-0.16}$ & 1.26&& 1.97$^{+0.45}_{-0.43}$ & 1.16$^{+0.14}_{-0.16}$ & 1.77$^{+0.05}_{-0.05}$ & 1.53$^{+0.44}_{-0.24}$ & 1.22 \\ \smallskip 
 13--15 & 0.97$^{+0.04}_{-0.05}$ & 1.82$^{+0.03}_{-0.03}$ & 1.26$^{+0.11}_{-0.10}$ & 1.24&& 1.94$^{+0.43}_{-0.43}$ & 1.09$^{+0.15}_{-0.17}$ & 1.82$^{+0.04}_{-0.04}$ & 1.35$^{+0.24}_{-0.17}$ & 1.21 \\ \smallskip 
 16--18 & 1.00$^{+0.07}_{-0.08}$ & 1.76$^{+0.05}_{-0.05}$ & 1.14$^{+0.17}_{-0.13}$ & 1.10&& 1.62$^{+0.68}_{-0.58}$ & 1.01$^{+0.25}_{-0.25}$ & 1.76$^{+0.05}_{-0.05}$ & 1.14$^{+0.31}_{-0.15}$ & 1.12 \\ \smallskip 
 19--20 & 0.97$^{+0.05}_{-0.06}$ & 1.68$^{+0.04}_{-0.04}$ & 1.15$^{+0.12}_{-0.10}$ & 1.41&& 1.58$^{+0.49}_{-0.46}$ & 0.96$^{+0.18}_{-0.20}$ & 1.68$^{+0.04}_{-0.04}$ & 1.14$^{+0.17}_{-0.13}$ & 1.44 \\ \smallskip 
 21--23 & 0.94$^{+0.04}_{-0.04}$ & 1.73$^{+0.03}_{-0.03}$ & 1.28$^{+0.09}_{-0.08}$ & 1.84&& 1.89$^{+0.34}_{-0.33}$ & 1.04$^{+0.12}_{-0.12}$ & 1.73$^{+0.03}_{-0.03}$ & 1.34$^{+0.16}_{-0.12}$ & 1.81 \\ \smallskip 
 24--25 & 0.95$^{+0.06}_{-0.07}$ & 1.70$^{+0.04}_{-0.04}$ & 1.31$^{+0.17}_{-0.16}$ & 1.13&& 2.41$^{+0.34}_{-0.33}$ & 1.25$^{+0.08}_{-0.08}$ & 1.77$^{+0.07}_{-0.06}$ & 2.31$^{+0.31}_{-0.28}$ & 0.98 \\ \smallskip 
 26--27 & 0.97$^{+0.06}_{-0.06}$ & 1.76$^{+0.04}_{-0.03}$ & 1.28$^{+0.17}_{-0.18}$ & 1.14&& 2.32$^{+0.32}_{-0.33}$ & 1.27$^{+0.07}_{-0.09}$ & 1.80$^{+0.08}_{-0.07}$ & 2.30$^{+0.31}_{-0.28}$ & 0.99 \\ \smallskip 
 28--30 & 1.09$^{+0.06}_{-0.07}$ & 1.77$^{+0.04}_{-0.04}$ & 1.33$^{+0.26}_{-0.16}$ & 1.24&& 2.06$^{+0.50}_{-0.57}$ & 1.25$^{+0.18}_{-0.22}$ & 1.77$^{+0.05}_{-0.05}$ & 1.54$^{+0.92}_{-0.35}$ & 1.21 \\ \smallskip 
 31--32 & 1.18$^{+0.04}_{-0.05}$ & 2.04$^{+0.09}_{-0.09}$ & 2.34$^{+0.19}_{-0.27}$ & 0.94&& 2.20$^{+0.45}_{-0.43}$ & 1.32$^{+0.11}_{-0.11}$ & 2.06$^{+0.11}_{-0.10}$ & 2.43$^{+0.25}_{-0.28}$ & 0.80 \\ \smallskip 
  33-34 & 1.08$^{+0.09}_{-0.10}$ & 2.01$^{+0.08}_{-0.08}$ & 1.14$^{+0.16}_{-0.13}$ & 1.36&& 3.08$^{+0.55}_{-0.53}$ & 1.64$^{+0.13}_{-0.13}$ & 2.28$^{+0.27}_{-0.21}$ & 3.12$^{+0.61}_{-0.80}$ & 1.30 \\ \smallskip 
 35--37 & 1.22$^{+0.06}_{-0.07}$ & 1.99$^{+0.05}_{-0.05}$ & 1.23$^{+0.17}_{-0.15}$ & 1.20&& 2.56$^{+0.38}_{-0.38}$ & 1.58$^{+0.10}_{-0.10}$ & 2.15$^{+0.13}_{-0.12}$ & 2.64$^{+0.39}_{-0.50}$ & 1.06 \\ \smallskip 
 38--40 & 1.07$^{+0.06}_{-0.07}$ & 1.90$^{+0.05}_{-0.05}$ & 1.11$^{+0.11}_{-0.11}$ & 1.34&& 2.82$^{+0.34}_{-0.37}$ & 1.55$^{+0.08}_{-0.10}$ & 2.06$^{+0.12}_{-0.14}$ & 2.83$^{+0.39}_{-0.68}$ & 1.15 \\ \smallskip 
 41--42 & 1.11$^{+0.07}_{-0.08}$ & 1.85$^{+0.06}_{-0.06}$ & 1.25$^{+0.22}_{-0.19}$ & 0.74&& 1.71$^{+0.62}_{-0.61}$ & 1.14$^{+0.22}_{-0.27}$ & 1.85$^{+0.07}_{-0.06}$ & 1.28$^{+0.39}_{-0.26}$ & 0.75 \\ 
\cutinhead{1998 April 21$^\mathrm{st}$}
   1--2 & 0.86$^{+0.06}_{-0.06}$ & 1.57$^{+0.05}_{-0.05}$ & 1.23$^{+0.14}_{-0.12}$ & 0.99&& 2.01$^{+0.66}_{-0.49}$ & 0.99$^{+0.24}_{-0.18}$ & 1.57$^{+0.06}_{-0.05}$ & 1.32$^{+0.94}_{-0.19}$ & 0.96 \\ \smallskip 
      3 & 0.91$^{+0.05}_{-0.06}$ & 1.55$^{+0.05}_{-0.05}$ & 1.53$^{+0.22}_{-0.23}$ & 1.07&& 1.71$^{+0.53}_{-0.54}$ & 0.94$^{+0.17}_{-0.22}$ & 1.55$^{+0.06}_{-0.06}$ & 1.56$^{+0.66}_{-0.36}$ & 1.08 \\ \smallskip 
      4 & 0.84$^{+0.05}_{-0.04}$ & 1.52$^{+0.04}_{-0.04}$ & 1.56$^{+0.29}_{-0.17}$ & 1.43&& 2.20$^{+0.27}_{-0.28}$ & 1.03$^{+0.07}_{-0.08}$ & 1.56$^{+0.07}_{-0.06}$ & 2.13$^{+0.32}_{-0.36}$ & 1.15 \\ \smallskip 
      5 & 0.84$^{+0.04}_{-0.04}$ & 1.47$^{+0.04}_{-0.04}$ & 1.39$^{+0.16}_{-0.12}$ & 1.35&& 2.29$^{+0.25}_{-0.25}$ & 1.08$^{+0.06}_{-0.06}$ & 1.56$^{+0.08}_{-0.12}$ & 2.47$^{+0.39}_{-0.97}$ & 1.32 \\ \smallskip 
      6 & 0.82$^{+0.05}_{-0.06}$ & 1.42$^{+0.04}_{-0.04}$ & 1.21$^{+0.21}_{-0.15}$ & 1.55&& 2.34$^{+0.28}_{-0.29}$ & 1.10$^{+0.07}_{-0.09}$ & 1.47$^{+0.06}_{-0.05}$ & 2.01$^{+0.42}_{-0.42}$ & 1.28 \\ \smallskip 
      7 & 0.82$^{+0.05}_{-0.05}$ & 1.45$^{+0.04}_{-0.04}$ & 1.28$^{+0.15}_{-0.12}$ & 1.05&& 2.02$^{+0.41}_{-0.43}$ & 0.97$^{+0.13}_{-0.16}$ & 1.46$^{+0.05}_{-0.05}$ & 1.46$^{+0.46}_{-0.26}$ & 1.00 \\ \smallskip 
   8--9 & 0.88$^{+0.03}_{-0.04}$ & 1.44$^{+0.03}_{-0.03}$ & 1.27$^{+0.11}_{-0.11}$ & 1.71&& 2.00$^{+0.34}_{-0.27}$ & 1.01$^{+0.13}_{-0.10}$ & 1.44$^{+0.04}_{-0.04}$ & 1.41$^{+0.75}_{-0.17}$ & 1.57 \\ \smallskip 
 10--12 & 0.90$^{+0.03}_{-0.03}$ & 1.47$^{+0.03}_{-0.03}$ & 1.34$^{+0.10}_{-0.09}$ & 0.96&& 1.80$^{+0.24}_{-0.25}$ & 0.96$^{+0.08}_{-0.09}$ & 1.47$^{+0.03}_{-0.03}$ & 1.41$^{+0.18}_{-0.14}$ & 0.93 \\ \smallskip 
 13--16 & 0.95$^{+0.03}_{-0.03}$ & 1.52$^{+0.03}_{-0.03}$ & 1.46$^{+0.16}_{-0.14}$ & 1.38&& 2.25$^{+0.21}_{-0.20}$ & 1.16$^{+0.05}_{-0.05}$ & 1.58$^{+0.06}_{-0.06}$ & 2.35$^{+0.27}_{-0.34}$ & 1.01 \\ 
\cutinhead{1998 April 23$^\mathrm{rd}$}
   1--5 & 0.96$^{+0.03}_{-0.03}$ & 1.58$^{+0.03}_{-0.03}$ & 1.40$^{+0.12}_{-0.10}$ & 1.76&& 2.09$^{+0.32}_{-0.25}$ & 1.13$^{+0.10}_{-0.09}$ & 1.59$^{+0.07}_{-0.03}$ & 1.70$^{+0.67}_{-0.25}$ & 1.50 \\ \smallskip 
  6--11 & 0.98$^{+0.02}_{-0.02}$ & 1.67$^{+0.02}_{-0.02}$ & 1.30$^{+0.07}_{-0.07}$ & 2.45&& 1.92$^{+0.21}_{-0.21}$ & 1.09$^{+0.08}_{-0.08}$ & 1.68$^{+0.03}_{-0.03}$ & 1.43$^{+0.19}_{-0.13}$ & 2.31 \\ \smallskip 
 12--19 & 1.07$^{+0.02}_{-0.02}$ & 1.69$^{+0.03}_{-0.03}$ & 1.46$^{+0.09}_{-0.09}$ & 1.79&& 2.02$^{+0.18}_{-0.18}$ & 1.20$^{+0.06}_{-0.06}$ & 1.69$^{+0.03}_{-0.03}$ & 1.62$^{+0.19}_{-0.14}$ & 1.37 \\
\enddata
\tablenotetext{a}{\footnotesize 
The Galactic hydrogen equivalent absorbing column is N$_\mathrm{H} = (1.61
\pm 0.1) \times 10^{20}$ cm$^{-2}$ (Lockman \& Savage~\citealp{lockman_savage95}).
The quoted errors are for 1 $\sigma$ for 3 parameters (i.e. $\Delta\chi^2 = 3.53$).
The number of degrees of freedom is 57.}
\tablenotetext{b}{\footnotesize 
The quoted errors are for 1 $\sigma$ for 4 parameters (i.e. $\Delta\chi^2 = 4.72$).
The number of degrees of freedom is 56.}
\end{deluxetable}